\documentclass[preprint,12pt]{emulateapj}

\usepackage{apjfonts}
\usepackage{natbib}
\usepackage{amssymb,amsmath}
\usepackage{graphicx}
\usepackage[colorlinks,urlcolor=red,citecolor=blue,linkcolor=cyan]{hyperref} 

\def\ltsima{$\; \buildrel < \over \sim \;$}
\def\lsim{\lower.5ex\hbox{\ltsima}}
\def\gtsima{$\; \buildrel > \over \sim \;$}
\def\gsim{\lower.5ex\hbox{\gtsima}}

\shortauthors{Albrecht et al.} \shorttitle{Obliquities of hot
  Jupiter host stars}
\begin{document}

\title{Obliquities of Hot Jupiter host stars:\\Evidence for tidal
  interactions and primordial misalignments}

\author{
  Simon Albrecht\altaffilmark{1}, 
  Joshua N.\ Winn\altaffilmark{1},
  John A.\ Johnson\altaffilmark{2},
  Andrew W.\ Howard\altaffilmark{3},
  Geoffrey W.\ Marcy\altaffilmark{3},\\
  R.\ Paul Butler\altaffilmark{4},
  Pamela Arriagada\altaffilmark{5},
  Jeffrey D.\ Crane\altaffilmark{6},
  Stephen A.\ Shectman\altaffilmark{6},
  Ian B.\ Thompson\altaffilmark{6},\\
  Teruyuki Hirano\altaffilmark{7},
  Gaspar Bakos,\altaffilmark{8,9},
  Joel D.\ Hartman\altaffilmark{8}
}

\altaffiltext{1}{Department of Physics, and Kavli Institute for
  Astrophysics and Space Research, Massachusetts Institute of
  Technology, Cambridge, MA 02139, USA}

\altaffiltext{2}{California Institute of Technology, Department of
  Astrophysics, MC249-17, Pasadena, CA 91125; NASA Exoplanet Science
  Institute (NExScI), USA}

\altaffiltext{3}{Department of Astronomy, University of California,
   Berkeley, CA 94720}

\altaffiltext{4}{Department of Terrestrial Magnetism, Carnegie
  Institution of Washington, 5241 Broad Branch Road NW, Washington, DCg
  20015, USA}
 
\altaffiltext{5}{Department of Astronomy, Pontificia Universidad
  Cat\'olica de Chile, Casilla 306, Santiago 22, Chile}

\altaffiltext{6}{The Observatories of the Carnegie Institution of
 Washington, 813 Santa Barbara Street, Pasadena, CA 91101, USA}

\altaffiltext{7}{Department of Physics, The University of Tokyo, Tokyo
  113-0033, Japan}
 
\altaffiltext{8}{Department of Astrophysical Sciences, Princeton
  University, Peyton Hall, Princeton, NJ 08544}

\altaffiltext{9}{Alfred. P. Sloan Fellow}

\altaffiltext{$\star$}{The data presented herein were collected with
  the the Magellan (Clay) Telescope located at Las Campanas
  Observatory, Chile; and the Keck~I telescope at the W.M.\ Keck
  Observatory, which is operated as a scientific partnership among the
  California Institute of Technology, the University of California and
  the National Aeronautics and Space Administration. }

\begin{abstract}

  We provide evidence that the obliquities of stars with close-in
  giant planets were initially nearly random, and that the low
  obliquities that are often observed are a consequence of star-planet
  tidal interactions. The evidence is based on 14 new measurements of
  the Rossiter-McLaughlin effect (for the systems HAT-P-6, HAT-P-7,
  HAT-P-16, HAT-P-24, HAT-P-32, HAT-P-34, WASP-12, WASP-16, WASP-18,
  WASP-19, WASP-26, WASP-31, Gl\,436, and Kepler-8), as well as a
  critical review of previous observations. The low-obliquity
  (well-aligned) systems are those for which the expected tidal
  timescale is short, and likewise the high-obliquity (misaligned and
  retrograde) systems are those for which the expected timescale is
  long.  At face value, this finding indicates that the origin of hot
  Jupiters involves dynamical interactions like planet-planet
  interactions or the Kozai effect that tilt their orbits, rather than
  inspiraling due to interaction with a protoplanetary disk. We
  discuss the status of this hypothesis and the observations that are
  needed for a more definitive conclusion.

\end{abstract}


\section{Introduction}
\label{sec:intro}

\setcounter{footnote}{0}

Exoplanetary science has been full of surprises. One of the biggest
surprises emerged at the dawn of this field: the existence of ``hot
Jupiters'' having orbital distances much smaller than an astronomical
unit (AU). It is thought that giant planets can only form at distances
of several AU from their host stars, where the environment is cooler
and solid particles are more abundant, facilitating the growth of
rocky cores that can then attract gaseous envelopes from the
protoplanetary disk.

Different mechanisms have been proposed which might transport giant
planets from their presumed birthplaces inward to where we find them.
Among the differences between the proposed mechanisms is that some of
them would alter the planet's orbital orientation and thereby change
the relative orientation between the stellar and orbital spin
\citep{nagasawa2008,fabrycky2007,naoz2011}, while others would preserve the
relative orientation \citep{lin1996}, or even reduce any primordial
misalignment \citep{cresswell2007}. For this reason, measuring the
stellar obliquity---the angle between stellar and orbital axes---has
attracted attention as a possible means of distinguishing between
different theories for the origin of hot Jupiters.

The stellar obliquity is an elusive parameter because the stellar
surface needs to be at least partially resolved by the observer. If
the system exhibits eclipses or transits, then it is possible to
detect anomalies in the spectral absorption lines of the eclipsed
star, which have their origin in the partial blockage of the rotating
photosphere. The precise manifestation of the rotation anomaly depends
on the angle between the projections of the stellar rotation axis and
orbital axis of the occulting companion. Credit for the first
definitive measurements has been apportioned to \cite{rossiter1924}
and \cite{mclaughlin1924}, after whom the effect is now
named. \cite{queloz2000} were the first to measure the
Rossiter-McLaughlin (RM) effect for a planet-hosting star, finding a
low obliquity. Since then many systems have been studied, including
those with misaligned stars \citep{hebrard2008}, retrograde orbits
\citep{winn2009,triaud2010}, a hot Neptune
\citep{winn2010c,hirano2011} and even a circumbinary planet
\citep{winn2011b}.

\cite{winn2010} found a possible pattern in the hot-Jupiter data,
namely, host stars hotter than $T_{\rm eff} \approx 6250$~K tend to
have high obliquities, while cooler stars have low
obliquities. \cite{schlaufman2010}, using a different method, also
found a preponderance of high obliquities among hot stars.
\cite{winn2010} speculated that this pattern was due to tidal
interactions.  Specifically, it was hypothesized that hot Jupiters are
transported inward by processes that perturb orbital inclinations and
lead to a very broad range of obliquities.  Cool stars ultimately
come into alignment with the orbits because they have higher rates of
tidal dissipation, due to their thick convective envelopes. Hot stars,
in contrast, lack thick convective envelopes and are unable to
reorient completely on Gyr timescales.

If this interpretation is correct, then not only should further
measurements be consistent with the hot/cool pattern, but also the
degree of alignment should be found to correlate with the orbital
period and planet-to-star mass ratio, parameters which also strongly
affect the rate of tidal dissipation. Since the study by
\cite{winn2010} the number of systems with RM measurements has nearly
doubled. Here we present the results of an additional 14 observations,
as well as a critical review of other published measurements
(including thorough re-analyses in three cases). We can now attempt a
comparison between the measured obliquities and the theoretical
distribution of obliquities that one would expect if tides were an
important factor. We also refer the reader to \cite{hansen2012}, who
recently performed a comparison with a similar motivation, without the
benefit of the new RM measurements presented here, but using more
sophisticated theoretical models and drawing qualitatively similar
conclusions.

This paper has two main parts. The first part is observational. We
describe our new observations of the RM effect
(Section~\ref{sec:obs}), our analysis method
(Section~\ref{sec:analysis}), and the details of individual systems
analyzed here (Section~\ref{sec:systems}). The second part
(Section~\ref{sec:discussion}) considers the distribution of
obliquities, seeks evidence for the expected signatures of tidal
effects, and considers the implications for the origin of hot
Jupiters. We summarize our results in Section~\ref{sec:summary}.

\section{Observations}
\label{sec:obs}

The 14 new observations presented in this paper were conducted with
the Keck\,I telescope and its High Resolution Spectrograph (HIRES;
\citealt{vogt1994}), for northern objects, and with the Magellan~II
(Clay) 6.5\,m telescope and the Planet Finder Spectrograph (PFS;
\citealt{crane2010}) for southern objects. Table \ref{tab:log} is a
log of the observations. To derive the relative RVs we compared the
spectra observed through the iodine cell with the stellar template
spectrum multiplied by an iodine template spectrum. The velocity shift
of the stellar template as well as the parameters of the point-spread
function (PSF) of the spectrograph are free parameters in this
comparison. The velocity shift of the template that gives the best fit
to an observed spectrum represents the measured relative RV for that
observation. In particular we used codes based on that of
\cite{butler1996}. The RVs used in this paper for all system are
presented in the Table~\ref{tab:rvs}.

\begin{table}
  \begin{center}
    \caption{Observation Log\label{tab:log}}
    \smallskip 
    \begin{tabular}{l c r  }
      \tableline\tableline
      \noalign{\smallskip}
         System & Observation night & Spectrograph \\
         & UT & \\
      \noalign{\smallskip}
      \hline
      \noalign{\smallskip}
      HAT-P-6  & 5/6 November 2010  & HIRES   \\
      HAT-P-7  & 23/24 July 2010 &  HIRES   \\
      HAT-P-16  & 26/27 December 2010  &    HIRES   \\
      HAT-P-24  & 27/28 September 2010 &   HIRES    \\
      HAT-P-32  & 5/6 December 2011  &  HIRES   \\
      HAT-P-34  & 2/3 September 2011 &   HIRES \\
      WASP-12  & 1/2 January 2012  &   HIRES \\
      WASP-16  & 3/4 July 2010  &  PFS   \\
      WASP-18  & 8/9 October 2011 &  PFS  \\
      WASP-19  & 20/21 May 2010 & PFS  \\
      WASP-26  & 17/18 August 2010 & PFS \\
      WASP-31  & 12/13 March 2012  &   HIRES \\
      Gl\,436  &   24/25 April 2010 & HIRES \\
      Kepler\,8  &  7/8 August 2011 &   HIRES   \\
     \noalign{\smallskip}
      \tableline
      \noalign{\smallskip}
      \noalign{\smallskip}
    \end{tabular}
 \end{center}
\end{table}

\section{Analysis of the Rossiter-McLaughlin effect}
\label{sec:analysis}

In the following sections we describe our strategy to obtain projected
obliquities from the observed radial velocities (RVs). Our goal was to
perform a relatively simple and homogeneous analysis and thereby
facilitate comparisons between systems in the sample. For some systems
for which we obtained new data, RM measurements have already been
published by other authors.  In order to obtain independent
measurements and compare the results, in most cases we did not include
the previous data in our analysis.

Observing the Rossiter-McLaughlin (RM) effect allows one to measure
the projected angle between the orbital and stellar spins ($\lambda$)
and the projected stellar rotation speed ($v \sin i_{\star}$). Here
$v$ indicates the stellar rotation speed and $i_{\star}$ the
inclination of the stellar spin axis towards the observer. Observing
the RM effect provides information on the transit geometry because the
distortion of the stellar absorption line---specifically its net
redshift or blueshift---depends on the RV of the hidden portion of the
stellar photosphere.

\begin{figure*}
  \begin{center}
    \includegraphics[width=14.5cm]{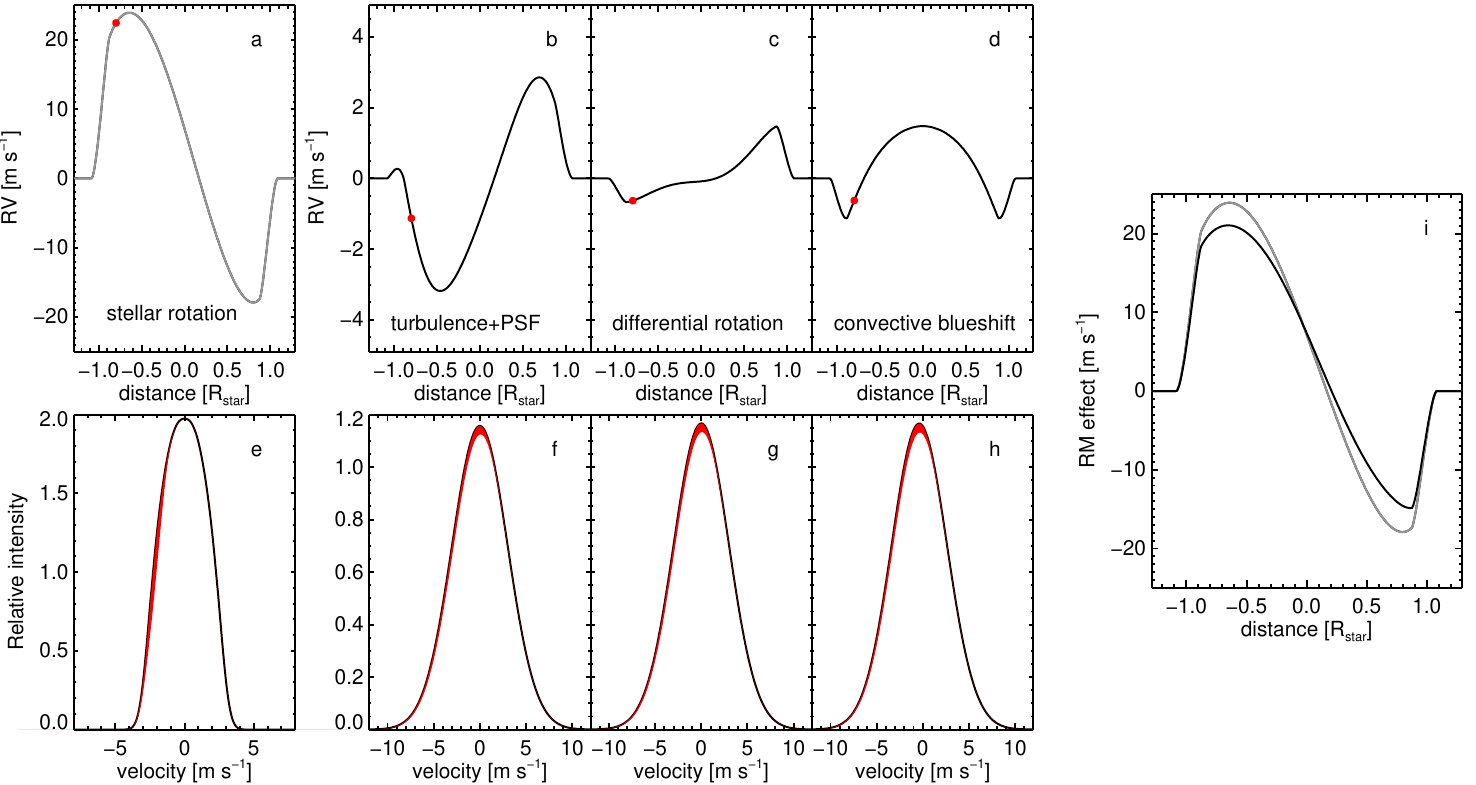}
    \caption {\label{fig:show_lines} {\bf Illustration of the
        influence of rotation, PSF broadening and macroturbulence,
        differential rotation, and convective blueshift on the RM
        effect.}   (a) The effect of rotation. The chosen parameters are $\lambda
      =40^{\circ}$, $v \sin i_{\star}= 3$~km\,s$^{-1}$, $R_{\rm
        p}/R_{\star}=0.12$ and an impact parameter of $0.2$. The
      $x$-coordinate is the planet's location relative to its
      position at inferior conjunction. (b) The effects of
      instrumental broadening
      ($2.2$~km\,s$^{-1}$) and macroturbulence ($\zeta_{\rm
        RT} = 3$~km\,s$^{-1}$). (c) The effect of
      solar-like differential rotation. (d) The effect of
      a solar-like convective blueshift. (e-h) The corresponding models for the stellar
      absorption lines. The red region correspond to the light lost
      during the particular transit phase indicated by a red dot in the upper
      panel. For the lower panels the radius of
      the planet was doubled, for improved visibility of the missing
      velocity components. (i) The combined model including all
      aforementioned effects. The gray line is the model
      from panel (a). }
  \end{center}
\end{figure*}

For example if the obliquity were low, then the planet would begin the
transit on the approaching half of the photosphere. The blockage
causes the absorption lines to appear slightly redshifted. During the
second half of the transit, the reverse would be true: the anomalous
RV would be a blueshift. In contrast, if the planet's trajectory were
entirely over the redshifted half of the star, then the RM effect
would be a blueshift throughout the transit. For a retrograde
configuration, the RM effect would first lead to a blueshift and then
a redshift.

In sections~\ref{sec:broaden}--\ref{sec:rv_var} we discuss some
specific effects which influence the RM signal, and how we included
these in our modeling. Section~\ref{sec:quantitative} gives the
details of our quantitative model of the RM effect, and the prior
information about each system that is used in the analysis.

\subsection{Line broadening}
\label{sec:broaden}

In addition to stellar rotation, microturbulence and macroturbulence
will affect stellar absorption lines and should be considered in the
description of the RM effect. Microturbulence is well described by a
convolution of the rotationally-broadened line with a Gaussian
function. Describing macroturbulence is more complex, as it depends on
the angle between the line of sight and the local surface normal. Near
the center of the stellar disk, the Doppler shifts are produced by
vertical (``radial'') motions, while near the limb, the Doppler shifts
are produced by motions along the stellar surface (``tangential''
motions). The net effect is obtained by spatial integration over the
visible hemisphere. \cite{hirano2011b} provided a semi-analytic
description of this effect, which we used in our model. The
macroturbulent field is parameterized by $\zeta_{\rm RT}$, the average
value of the radial and tangential velocities (assumed to be equal).

What value of $\zeta_{\rm RT}$ should be used to model the RM effect
for a given system? Measuring macroturbulence from the line profile is
challenging, as its measurement is correlated with the measurement of
$v \sin i_{\star}$. In principle the RM effect itself could be used to
measure macroturbulence. However for most of the systems analyzed here
the data sets are not precise enough to allow $\zeta_{\rm RT}$ to be a
free parameter. If the macroturbulence cannot be determined by the RM
effect, one might think its value also does not matter for the
analysis.  However, as described below, we used prior information on
$v \sin i_{\star}$ as a constraint on our model, and this information
can only be properly incorporated if macroturbulence is taken into
account. We therefore made use of the relation given by
\citep[][eq.\,4]{gray1984},
\begin{equation}
  \label{equ:macro}
  \zeta_{\rm RT} = 3.95\,T_{\rm eff}  - 19.25\,\, ,
\end{equation}
where $\zeta_{\rm RT}$ is in km\,s$^{-1}$ and $T_{\rm eff}$ in
thousands of Kelvin. \cite{gray1984} gives an uncertainty of
$0.2$~km\,s$^{-1}$ for this relationship between $T_{\rm eff}$ and
$\zeta_{\rm RT}$. We adopt a more conservative uncertainty estimate of
$1.5$~km\,s$^{-1}$.  We decided to use the relationship by
\cite{gray1984} for macroturbulence instead of the relationship by
\cite{valenti2005}, because the latter authors derived the
relationship between $T_{\rm eff}$ and $\zeta_{\rm RT}$ for F stars by
extrapolation from later spectral types, while \cite{gray1984}
obtained macroturbulence values for hotter stars directly. Many of the
stars in our sample have a $T_{\rm eff}$ greater then $6000$~K.

Additional contributions to the linewidth come from collisional
broadening, and Zeeman splitting, which are small compared to the
effects described above. We account for them, at least approximately,
by convolving the disk-integrated line profiles with a Lorentzian
function of width $1$~km\,s$^{-1}$.

\begin{table*}
  \begin{center}
    \caption{Prior knowledge\label{tab:priors}}
    \smallskip 
    \begin{tabular}{l r r@{$\pm$}l
        r@{$\pm$}l  r@{$\pm$}l 
        r@{$\pm$}l   r@{$\pm$}l 
        r@{$\pm$}l r@{$\pm$}l 
        r
          }
      \tableline\tableline
      \noalign{\smallskip}
         System & \multicolumn{1}{c}{Jitter} &\multicolumn{2}{c}{ $T_{c}- 2\,400\,000$} 
         & \multicolumn{2}{c}{  $T_{41}$}    &   \multicolumn{2}{c}{$T_{21}$}  
         & \multicolumn{2}{c}{ $R_{\rm p}/R_{\star}$}   & \multicolumn{2}{c}{$u1_{\rm}$ + $u2_{\rm}$} 
         &  \multicolumn{2}{c}{ $\zeta_{\rm RT}$}  &   \multicolumn{2}{c}{$v \sin i_{\star}$}     
         & ref. \\
         &  \multicolumn{1}{c}{(m\,s$^{-1}$)}  & \multicolumn{2}{c}{ (BJD$_{\rm TDB}$) } 
         & \multicolumn{2}{c}{ (hr)}  & \multicolumn{2}{c}{ (hr) }  
         & \multicolumn{2}{c}{     }   &  \multicolumn{2}{c}{     }     
         &  \multicolumn{2}{c}{(km\,s$^{-1}$)} & \multicolumn{2}{c}{(km\,s$^{-1}$)}  
         & \\
      \noalign{\smallskip}
      \hline
      \noalign{\smallskip}
      HAT-P-6  &   7.0  &     54035.67652&0.00196  & 3.506&0.041  & 0.451&0.053  & 0.0934&0.0005  & 0.67&0.10  & 5.52&1.50  &   8.7& 1.5 &  1    \\
      HAT-P-7  &   6.0  &     55401.91904&0.00002  & 3.941&0.002  & 0.373&0.004  & 0.0777&0.0001  & 0.70&0.10  & 5.18&1.50  &   3.8& 1.5 &  2    \\
      HAT-P-16  &   4.0  &     55027.59293&0.00574  & 3.062&0.031  & 0.360&0.067  & 0.1071&0.0014  & 0.71&0.10  & 4.88&1.50  &   3.5& 1.5 & 3      \\
      HAT-P-24  &  12.0  &     55216.97667&0.00071  & 3.694&0.019  & 0.338&0.029  & 0.0970&0.0012  & 0.70&0.10  & 5.22&1.50  &  10.0& 1.5 & 4      \\
      HAT-P-32  &  42.0  &     54420.44637&0.00069  & 3.108&0.007  & 0.413&0.010  & 0.1508&0.0004  & 0.69&0.10  & 4.96&1.50  &  20.7& 1.5 &  5     \\
      HAT-P-34  &  25.0  &     55431.59629&0.00126  & 3.492&0.038  & 0.290&0.062  & 0.0801&0.0026  & 0.69&0.10  & 5.32&1.50  &  24.0& 1.5 &  6     \\
      WASP-12  &   4.5  &     54508.97605&0.00146  & 3.001&0.037  & 0.324&0.040  & 0.1119&0.0020  & 0.71&0.10  & 5.10&1.50  &   2.2& 1.5 &  7,8     \\
      WASP-16  &  10.0  &     54584.42952&0.00509  & 1.934&0.031  & 0.504&0.096  & 0.1079&0.0012  & 0.74&0.10  & 4.18&1.50  &   3.0& 1.5 &  9     \\
      WASP-18  &   8.0  &     54221.48163&0.00155  & 2.191&0.012  & 0.218&0.034  & 0.0970&0.0010  & 0.68&0.10  & 5.26&1.50  &  11.0& 1.5 &   10    \\
      WASP-19  &  20.0  &     55168.96879&0.00011  & 1.572&0.007  & 0.324&0.024  & 0.1425&0.0014  & 0.76&0.10  & 3.87&1.50  &   4.0& 2.0  &  11,12   \\
      WASP-26  &  10.0  &     55228.38916&0.00075  & 2.383&0.043  &  0.057&0.125  & 0.1011&0.0017  & 0.74&0.10  & 4.56&1.50  &   2.4& 1.5  &   13,14  \\
      WASP-31  &   6.0  &    55209.71890&0.00280  & 0.110&0.001  & 0.029&0.004  & 0.1271&0.0011  & 0.69&0.10  & 5.11&1.50  &   7.9& 1.5  &  15 \\
      Gl\,436  &   1.5  &     54235.83624&0.01221  & 1.012&0.010  & 0.213&0.028  & 0.0825&0.0078  & 0.87&0.10  & 0.56&1.50  &   \multicolumn{2}{c}{$\,\,\,\,0+5$}  &  this work \\
      Kepler-8  &  10.0  &     55781.90611&0.00032  & 3.257&0.010  & 0.557&0.029  & 0.0948&0.0006  & 0.71&0.10  & 4.97&1.50  &  10.5& 1.5 &  16     \\
      HAT-P-2  &   0.0  &     54387.49375&0.00075  & 4.289&0.031  & 0.338&0.072  & 0.0723&0.0006  & 0.70&0.10  & 5.09&1.50  &  20.8& 1.5 & 17      \\
      HD\,149026  &   2.5  &     54456.78835&0.00080  & 3.230&0.150  & 0.227&0.024  & 0.0507&0.0009  & 0.72&0.10  & 4.89&1.50  &   6.0& 1.5 & 18,19      \\
      HD\,209458  &   2.0  &     51659.93742&0.00002  & 3.072&0.003  & 0.438&0.008  & 0.1211&0.0001  & 0.71&0.10  & 4.64&1.50  &   4.5& 1.5 &  20     \\
    \noalign{\smallskip}
      \tableline
      \noalign{\smallskip}
      \noalign{\smallskip}
    \end{tabular}
    \tablerefs{ (1) \cite{noyes2008}: (2) \cite{pal2008}; (3)\cite{buchhave2010} ;
      (4) \cite{kipping2010}; (5) \cite{hartman2011}; (6) \cite{bakos2012};
      (7) \cite{maciejewski2011}; (8) \cite{hebb2009}; (9)\cite{lister2009}; 
      (10) \cite{hellier2009}; (11) \cite{hebb2010b}; (12) \cite{hellier2011}; 
      (13) \cite{smalley2010}; (14) \cite{anderson2011}; (15) \cite{anderson2011c};
      (16) \cite{jenkins2010}; (17) \cite{pal2010};  (18) \cite{carter2009}; 
      (19) \cite{sato2005}; (20) \cite{laughlin2005} }
  \end{center}
\end{table*}

\subsection{Convective blueshift}
\label{sec:cb}

Upward-flowing material in a convective cell is hotter and more
luminous than downward-flowing material. This leads to a net blueshift
in the disk-integrated light, known as the convective blueshift (CB).
The CB is strongest in light received from the center of the disk, and
weaker near the stellar limb. Disk integration leads to an overall
Doppler shift and an asymmetry in the stellar absorption lines. The
overall Doppler shift is of order 1~km~s$^{-1}$, but this is
unimportant for our purposes, as we are only interested in relative
RVs. However, the motion of the transiting planet will cause time
variations in the net CB of the exposed surface of the star. The
time-varying component of the CB effect is of order 1~m~s$^{-1}$ for
transits of late-type stars (Figure~\ref{fig:show_lines}).  This is a
small effect but for completeness it was included in our model.

To describe the effect of the CB, we used a numerical model based on
work by \cite{shporer2011}. For the stars in our sample with $T_{\rm
  eff} < 6000$~K we assumed the net convective blueshift to be similar
to that of the Sun: an effective radial velocity of the photosphere of
$500$~m\,s$^{-1}$.  For hotter stars we assumed an effective radial
velocity of $1000$~m\,s$^{-1}$.  The single exception to this rule was
Gl\,436, which is much cooler than the other stars ($T_{\rm eff} =
3350\pm300$~K). For this system we assumed a smaller blueshift of
$200$~m\,s$^{-1}$. In principle one could estimate the overall CB for
each system based on the observed line bisectors, but we did not
pursue this approach because our first-order analysis revealed that
the CB effect is relatively unimportant.

\subsection{Differential rotation}
\label{sec:diff}

For this study we neglected the possible influence of differential
surface rotation on the RM effect. \cite{albrecht2012} concluded that
modeling differential rotation is justified only if the transit chord
spans a wide range of stellar latitudes, the RM effect is detected
with a very high signal-to-noise ratio, and excellent prior
constraints are available for the limb darkening (not only as it
affects the continuum but also the absorption lines). These conditions
are not met simultaneously for any of the systems in our sample. See
\cite{albrecht2012} for further discussion and an application of a
model with differential rotation to the WASP-7 system.

\subsection{RV measurements during transits}
\label{sec:rvs}

A model which aims to simulate RV measurements made during transits
needs to take into account that the measured RVs represent the output
of a Doppler-measuring code. These codes are complex but in essence
they locate the peak of a cross-correlation between a stellar template
spectrum obtained outside transit, and a spectrum observed during
transit. Depending on the system parameters these RVs can
significantly differ from RVs representing the first moment of the
absorption line. The later are often used in RM work potentially
leading to systematic errors in the derived parameters
\citep{hirano2010,hirano2011b}.

\cite{hirano2011b} presented a semi-analytic description of the shift
in the cross-correlation peak as a function of the transit parameters,
including the dependence of the RVs on the stellar rotation velocity
and obliquity, the microturbulent and macroturbulent velocities, the
differential rotation profile, and the PSF of the spectrograph. We
used this model, after extending it to include the convective
blueshift. We also tested the results of the code by
\cite{hirano2011b} with the fully numerical disk-integration code of
\cite{albrecht2007}, including the same physical effects, finding good
agreement. Figure~\ref{fig:show_lines} shows the influence of the
effects discussed in this and the previous sections on the stellar
absorption lines.

\subsection{Other RV variation sources}
\label{sec:rv_var}

In addition to the RM effect we must also model the changes in radial
velocity due to the star's orbital motion. Over the course of the
transit night, this motion can usually be represented as a linear
function of time, and parameterized by a constant acceleration. For
ease of comparison with existing orbital solutions, the out-of-transit
velocity trend may also be parameterized by the orbital velocity
semi-amplitude ($K_{\star}$), for fixed values of the other orbital
parameters (period $P$, eccentricity $e$, argument of pericenter
$\omega$).

The question arises whether to allow $K_{\star}$ to be a completely
free parameter in our analysis, or whether to use a prior constraint
based on a previously published orbital solution. The answer is not
obvious because stars exhibit intrinsic RV noise that may have
different amplitudes on different timescales, ranging from a few
minutes to days. Short-timescale noise will simply degrade our
measurement accuracy, but noise on timescales longer than $\sim$6~hr
(such as the noise produced by rotating starspots) will introduce
trends in the RV signal over the course of the transit night. Without
a specific model for the frequency content of the intrinsic RV noise,
it is difficult to combine the data obtained sporadically over days or
months with the data obtained with much finer time sampling on a
single night. Specifically, if we would choose to constrain $K_\star$
(and therefore the transit-night velocity gradient) based on the
orbital solution, there is a risk of introducing a bias in our results
because of actual transit-night velocity gradient could be affected by
starspot variability or other intrinsic RV noise with timescales
longer than a few hours \citep{bouchy2008,albrecht2011b,albrecht2012}.

\begin{table*}
  \begin{center}
    \caption{Results for rotation parameters \label{tab:results}}
    \smallskip 
    \begin{tabular}{l 
        r@{$\pm$}l  r@{$\pm$}l 
        r@{$\pm$}l  r@{$\pm$}l 
        r@{$\pm$}l |  r@{$\pm$}l 
        r@{$\pm$}l  r@{$\pm$}l 
        r  }
      \tableline\tableline
      \noalign{\smallskip}
      \multicolumn{11}{c}{Final results} & \multicolumn{7}{c}{ With prior on $K_{\star}$\footnote{The results in columns 8-9 are based
           on an analysis in which a prior constraint was imposed on
           $K_{\star}$, using the value reported in column 7. We
           prefer the values in columns 4-5 for reasons
           described in Section~\ref{sec:rv_var}.} }  \\
       \noalign{\smallskip}
         System 
         & \multicolumn{2}{c}{ $\sqrt{v \sin i_{\star}} \sin \lambda$}    & \multicolumn{2}{c}{ $\sqrt{v \sin i_{\star}} \cos \lambda$}  
         &  \multicolumn{2}{c}{  $v \sin i_{\star}$} &   \multicolumn{2}{c}{ $\lambda$} 
         & \multicolumn{2}{c}{$K_{\star}$ transit}      & \multicolumn{2}{c}{ $K_{\star}$ literature}
         &  \multicolumn{2}{c}{  $v \sin i_{\star}$} &     \multicolumn{2}{c}{ $\lambda_{\rm}$} 
        & Ref.\footnote{Reference for the value of $K_{\star}$ in column 7.}\\
         & \multicolumn{2}{c}{(km\,s$^{-1}$)} &\multicolumn{2}{c}{(km\,s$^{-1}$)}  
         & \multicolumn{2}{c}{(km\,s$^{-1}$)}  &\multicolumn{2}{c}{($^{\circ}$)}  
          & \multicolumn{2}{c}{(km\,s$^{-1}$)} &\multicolumn{2}{c}{(km\,s$^{-1}$)}  
         & \multicolumn{2}{c}{(km\,s$^{-1}$)} &\multicolumn{2}{c}{($^{\circ}$)}  
      &  \\
      \noalign{\smallskip}
      \hline
      \noalign{\smallskip}
      HAT-P-6      &   0.702&0.269  &   -2.689&0.131   &   7.8& 0.6  &  165&6  &        188&20 &  115.5 &4.2  &   7.0&0.6  & 175&4 & 1 \\
      HAT-P-7      &   0.695&0.361  &  -1.476&0.196   &   2.7& 0.5  & 155& 37
      \footnote{The internal uncertainty for the $\lambda$ measurement
        in this system is $14^{\circ}$. The larger uncertainty reported here
        is the standard deviation of the 3 independent measurements,
        as discussed in section ~\ref{sec:hatp7}.}  &        214&5 &  213.5&1.9    &   2.7&0.4  & 155&14 & 2 \\
      HAT-P-16    &  -0.051&1.413 &  1.327&0.364  &   3.1& 1.0  & \multicolumn{2}{c}{$-2^{+55}_{-46}$}  &        536&60 &  531.1&2.8  &   2.7&0.8  & -6&37 & 3 \\
      HAT-P-24    &   1.158&0.851  &  3.087&0.297  & 11.2& 0.9  &   20& 16 &        129&41 &   83&3.4     &   11.0&0.9  & 14&16 & 4 \\
      HAT-P-32    &   4.517&0.168  &  0.396&0.117  & 20.6& 1.5  &   85.0& 1.5 &       77&26 &  122.8&23.2   &   20.6&1.5  & 85.3&1.5 & 5  \\
      HAT-P-34    &  -0.004&1.164 &  4.822&0.179  & 24.3& 1.2  &     0& 14 &        421&32 &  343.1&21.3    &   24.6&1.2  & -7&12 & 6 \\
      WASP-12      &   1.044&0.399  &  0.640&0.210  &  \multicolumn{2}{c}{$1.6^{+0.8}_{-0.4}$}  & \multicolumn{2}{c}{$59^{+15}_{-20}$} &  204.4&2.4 &  226&4    &  \multicolumn{2}{c}{$1.4^{+0.9}_{-0.5}$}   & \multicolumn{2}{c}{$63^{+14}_{-21}$} & 7    \\
      WASP-16      &   0.324&0.617  &  1.674&0.376  &   3.2& 0.9  & \multicolumn{2}{c}{$11^{+26}_{-19}$}   &        353&54 &  116.7&2.4    &   5.4&0.7  & \multicolumn{2}{c}{$-36^{+7}_{-10}$} & 8 \\
      WASP-18      &   0.774&0.421  &  3.234&0.126  & 11.2& 0.6  &   13&  7  &        1768&5 &   1818.3&8.0   &   10.9&0.6  & 12.6&7 & 9 \\
      WASP-19      &   0.552&0.371  &  1.984&0.241  &   4.4& 0.9  &   15& 11 &        200&17 &  256&5     &   3.7&0.9  & 7&13 & 10 \\
      WASP-26      &  -0.803&0.722 &  1.117&0.304  &   2.2& 0.7  &    \multicolumn{2}{c}{$-34^{+36}_{-26}$}   &        137&15 &  137.7&1.5    &   2.6&0.9  &\multicolumn{2}{c}{$-42^{+34}_{-25}$} & 11 \\
      WASP-31      &   0.076&0.130  &   2.605&0.121  &  6.8& 0.6  &    -6&  6  &        53&19 &  58.1&3.4   &   7.2&0.6  & -6&3 & 12 \\
      Gl\,436         &  -0.081&0.357 &   -0.192&0.584   &    \multicolumn{2}{c}{$<0.4$}  &   \multicolumn{2}{c}{--}  &        19&3 &   18.34&0.52      &     \multicolumn{2}{c}{$<0.4$}  &  \multicolumn{2}{c}{--}  & 13 \\
      Kepler-8      &  0.255&0.353 &  2.960&0.179  &   8.9& 1.0  &        5&  7  &        74&37 & 68.4&12.0 &   8.8&0.9 & -5&6 & 14 \\
      HAT-P-2      &   0.697&0.341  &  4.349&0.101  & 19.5& 1.4  &     9&  10 &        808&9 &   983.9&17.2   &   19.2&0.7  & 9&4 & 15 \\
      HD\,149026 &   0.558&0.339  &  2.706&0.157  &   7.7& 0.8  &   12&  7  &        31&6 &  43.3&1.2   &  7.2&0.7  & 4&6  & 16\\
      HD\,209458 &  -0.167&0.237 &  2.084&0.056  &   4.4& 0.2  &   -5& 7 &        128&32 &   84.67&0.70     &   4.5&0.2  & 3&3 & 17  \\
    \noalign{\smallskip}
      \tableline
      \noalign{\smallskip}
      \noalign{\smallskip}
    \end{tabular}
  \end{center}
 \tablerefs{ (1) \cite{noyes2008}: (2) \cite{pal2008}; (3)\cite{buchhave2010} ;
      (4) \cite{kipping2010}; (5) \cite{hartman2011}; (6) \cite{bakos2012};
      (7) \cite{hebb2009}; (8)\cite{lister2009}; 
      (9) \cite{hellier2009}; (10) \cite{hebb2010b}; (11) \cite{anderson2011}; 
      (12) \cite{anderson2011c}; (13) \cite{maness2007}
     (14) \cite{jenkins2010}; (15) \cite{pal2010};    (16) \cite{sato2005}; (17) \cite{torres2008} }
\end{table*}

One might proceed by downweighting the constraint on $K_\star$ by some
amount deemed to be appropriate, or fitting all of the RV data using a
frequency-dependent noise model. For this work we placed a high
premium on simplicity and homogeneity, and decided to decouple the
transit-night data from the rest of the data. We allowed $K_\star$ to
be a completely free parameter and determine the velocity gradient
from the transit-night data alone. This was done for all the systems
in our sample, even those where we find no obvious sign of intrinsic
RV noise. This leads to larger uncertainties, and for some systems we
find wider confidence intervals for $\lambda$ than other researchers
using similar data or even noisier data. While in some systems this
might be an unnecessary precaution, the stars' characteristics are in
general not well enough known to make a principled case-by-case
decision. Table~\ref{tab:results} gives the values of $K_{\star}$ from
published orbital solutions, along with the the values of $K_{\star}$
derived from our model which was fitted to the transit-night
data.

\begin{figure}
  \begin{center}
  \includegraphics{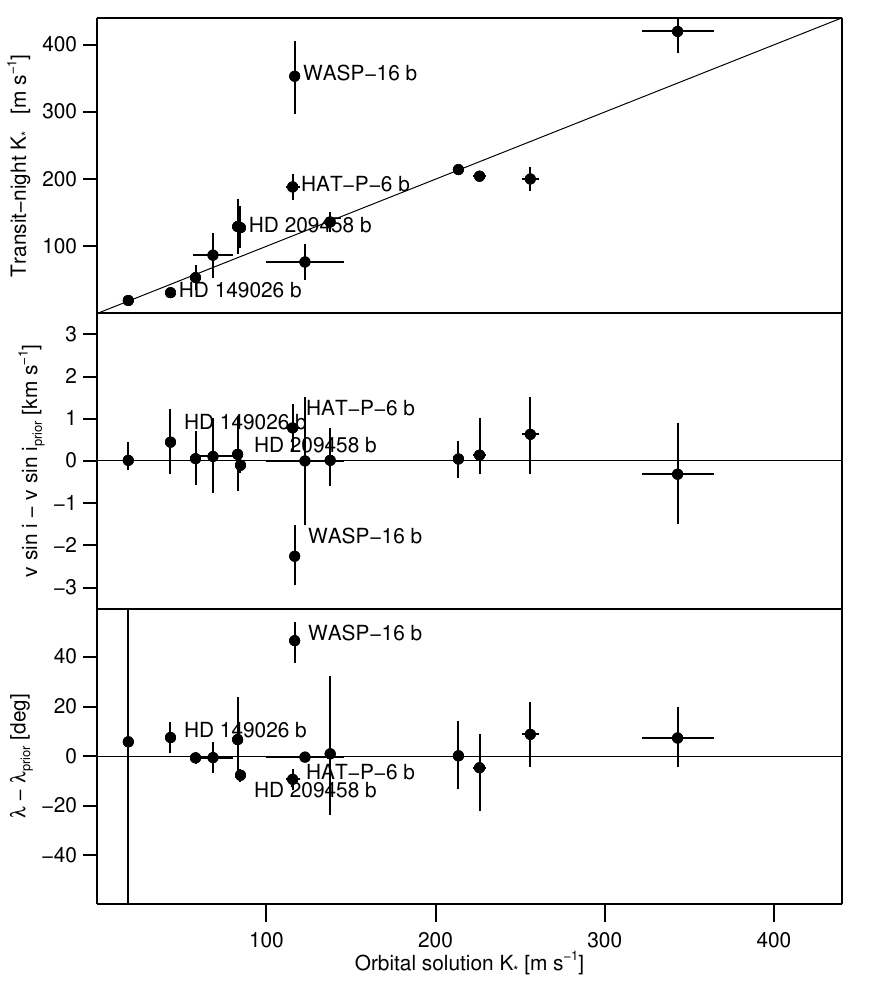}
  \caption {\label{fig:k1prior} {\bf Comparison between results with
      and without a prior on $K_{\star}$.}  {\it Upper panel:}
    Measurements of $K_{\star}$ based on the transit-night velocity
    gradient,
    versus previously reported measurements of $K_{\star}$ from
    orbital solutions based on RV data obtained over months or years.
    (The three systems with the
    largest $K_{\star}$ are omitted to permit the other systems to be
    viewed more clearly.) While for many systems the
    results are consistent, there are a number of
    systems showing disagreement, possibly due to stellar RV noise with a timescale of hours to days.
    {\it Middle panel:} Difference between $v \sin i_{\star}$ values
    obtained with and without a constraint on $K_\star$ based on the
    orbital solution. {\it Lower panel:} Same, but for $\lambda$.}
  \end{center}
\end{figure}

For those readers who do not agree with our reasoning on this point or
who are simply curious about the effect of imposing a prior on
$K_\star$, Table~\ref{tab:results} and Figure~\ref{fig:k1prior} also
provide the results for $\lambda$ and $v\sin i_\star$ obtained by
imposing a prior on $K_{\star}$ based on the orbital solution.

\subsection{Quantitative Analysis}
\label{sec:quantitative}

Having described the model, we now describe the procedure for
parameter estimation.  For each observed transit, we calculated the
expected time of inferior conjunction ($T_{\rm c}$) based on the
published ephemerides.  This predicted time was used as a prior in our
analysis of the transit-night RV data. The out-of-transit variation
was parameterized by the semi-amplitude of the projected stellar reflex
motion ($K_{\star}$), and an arbitrary additive constant velocity
($\gamma$).

The transit geometry was described by the following parameters: the
radius of the star in units of the orbital semi-major axis
($R_{\star}/a$), the radius of the planet in units of the stellar
radius ($R_{\rm p}/R_{\star}$), and the cosine of the orbital
inclination ($\cos i_{\rm o}$). We imposed Gaussian priors on $R_{\rm
  p}/R_{\star}$, the total transit duration from first to last contact
($T_{41}$), and the ingress duration ($T_{21}$), based on previously
reported photometric analyses. By choosing those particular parameters
we minimized the correlations among their uncertainties
\citep{carter2008}, consistent with our treatment of the priors as
independent Gaussian functions. However, small correlations do exist
and affect $T_{21}$ most strongly. In the interest of homogeneity we took
the simple and conservative approach of doubling the reported
uncertainty in $T_{21}$ for all systems. Table~\ref{tab:priors}
specifies all the priors that were used in practice, including the
increased uncertainty in $T_{21}$.

For systems where values for $T_{41}$ or $T_{21}$ were not previously
reported in the literature we obtained a transit light curve and
derived those parameters ourselves, using the transit model of
\cite{mandel2002}. The uncertainty intervals in $R_{\rm p}/R_{\star}$,
$T_{41}$, and $T_{21}$ were then used as Gaussian priors in the RV
analysis in the same way as for all the other systems, including the
doubled uncertainty in $T_{21}$.

We assumed a quadratic limb-darkening law with parameters $u_{1 \rm
  RM}$ and $u_{2 \rm RM}$ selected from the tables by
\cite{claret2000} for the Johnson $V$ band (similar to the iodine
spectral region). We used the ``jktld'' \footnote{This code is kindly
  made available by J.\ Southworth: {\tt
    http://www.astro.keele.ac.uk/jkt/codes/jktld.html}} tool to query
the ATLAS models for all systems except Gl\,436, for which we used the
tables based on the Phoenix code. We placed a Gaussian prior on $u_{1
  \rm RM}+u_{2 \rm RM}$ with a width of $0.1$. The difference $u_{1
  \rm RM}-u_{2 \rm RM}$ was held fixed at the tabulated value, since
this combination is weakly constrained by the data and has minimal
effect on the other parameters.

Because $\lambda$ and $v\sin i_\star$ have correlated uncertainties we
used the fitting parameters $\sqrt{v\sin i_\star} \cos\lambda$ and
$\sqrt{v\sin i_\star} \sin\lambda$. A final constraint used in our
fitting statistic is the prior knowledge concerning the projected
rotation speed, which was measured for all systems in our sample. Here
we adopted a minimum uncertainty of $1.5$~km\,s$^{-1}$ even when the
reported uncertainty was lower. The $v\sin i_\star$ measurements in
our sample were conducted by different researchers using different
approaches and it is not clear that the scales for the different
approaches are identical, particularly for the very challenging cases
of low projected rotation speeds. Furthermore we expect that our lack
of knowledge on differential rotation, macroturbulence, and
spectroscopic limb darkening introduces systematic errors at that
level. All constraints used in our fits are listed in
table~\ref{tab:priors}.

Finally, we must specify the width of a Gaussian function representing
the contribution to the line width due to both microturbulence and the
PSF of the spectrograph. The influence of both can be approximated by
a convolution with a single Gaussian, despite their difference in
origin \citep{hirano2011b}. Assuming a value of $2$~km\,s$^{-1}$ for
the microturbulence parameter and $2.2$~km\,s$^{-1}$ for the PSF width
at $5500$~\AA{}, we obtain a $\sigma$ of $3$~km\,s$^{-1}$ for this
purpose. The results for $\lambda$ and $v\sin i_\star$ do
not depend strongly on the values of these parameters.

To derive confidence intervals for the parameters we used the Markov
Chain Monte Carlo algorithm. Before starting the chain we also added
``stellar jitter'' terms in quadrature to the internally-estimated
uncertainty of the RVs to obtain a reduced $\chi^{2}$ close to
unity. In making this step we assumed that the uncertainties in the RV
measurements within a given night are uncorrelated and
Gaussian. Table~\ref{tab:rvs} reports the original,
internally-estimated uncertainties without any jitter term. The added
terms are listed for each system in Table~\ref{tab:priors}

The results reported in Table~\ref{tab:results} are the median values
of the posterior distribution. The quoted uncertainty intervals
represent the range that excludes $15.85$\% of the values on each
side of the posterior distribution, and encompass $68.3$\% of the
probability.

\begin{figure*}
  \begin{center}
  \includegraphics{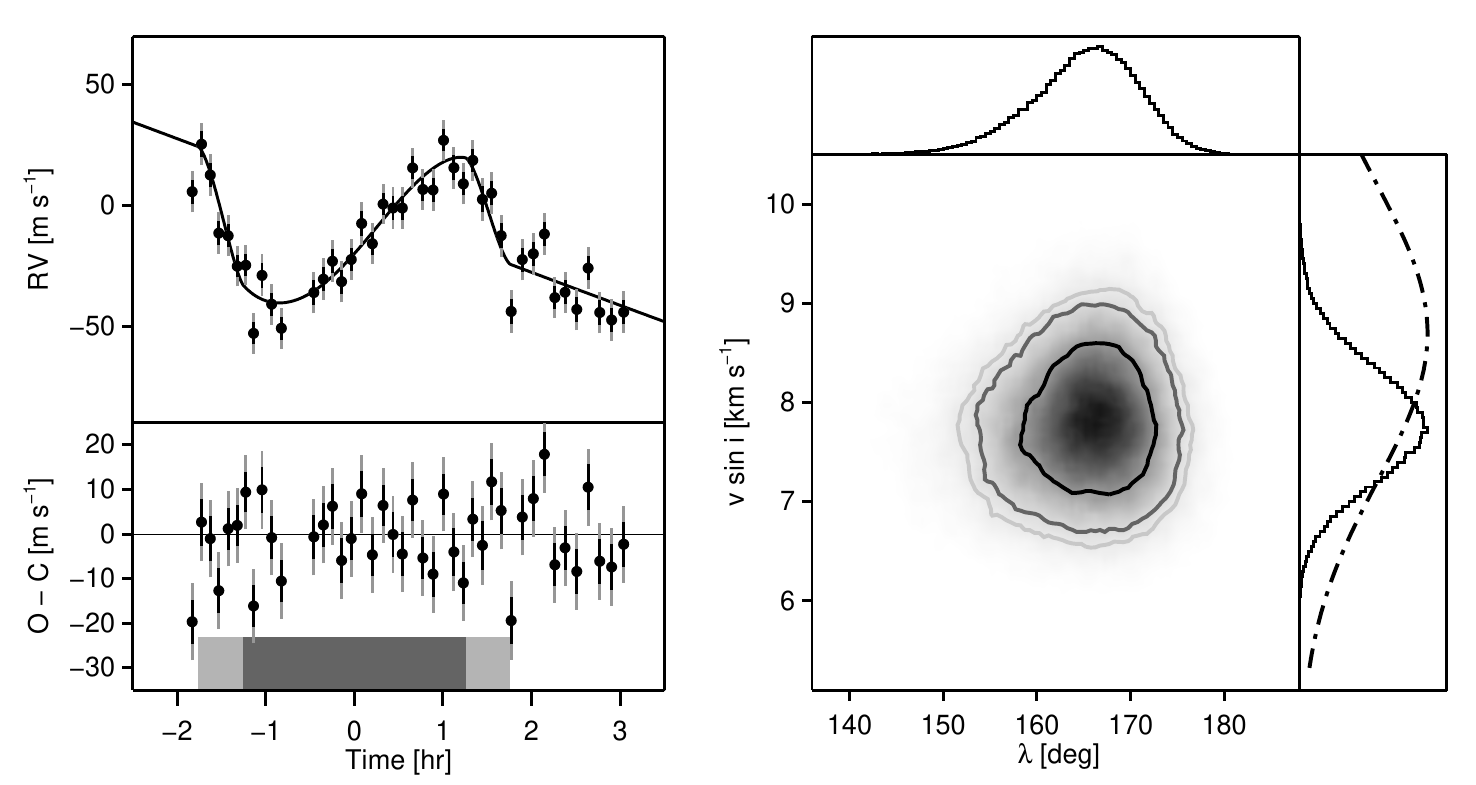}
  \caption {\label{fig:hatp6} {\bf Spectroscopy of HAT-P-6 transit}
    {\it Left panels--} The radial velocities measured before, during,
    and after transit are plotted as a function of time from inferior
    conjunction. The black error bars indicate the internal RV
    uncertainties. The gray error bars also include ``stellar jitter''
    as explained in the text. The upper panel shows the measured RVs,
    with $\gamma$ subtracted, and the best-fitting model. The lower panel shows the RVs after subtracting the
    best-fitting orbital model. The light and dark gray bars in the
    lower panel indicate times of first, second, third, and fourth
    contact. {\it Right panels--} The gray scale indicates the
    posterior probability density, marginalized over all other
    parameters. The black, dark gray, and light gray contours
    represent the 2-D 68.3\%, 95\%, and 99.73\% confidence limits. The
    one-dimensional marginalized distributions are shown on the sides
    of the contour plot. The dash-dotted line shows the prior applied
    to $v\sin i_\star$.}
  \end{center}
\end{figure*}

\begin{figure*}
  \begin{center}
  \includegraphics{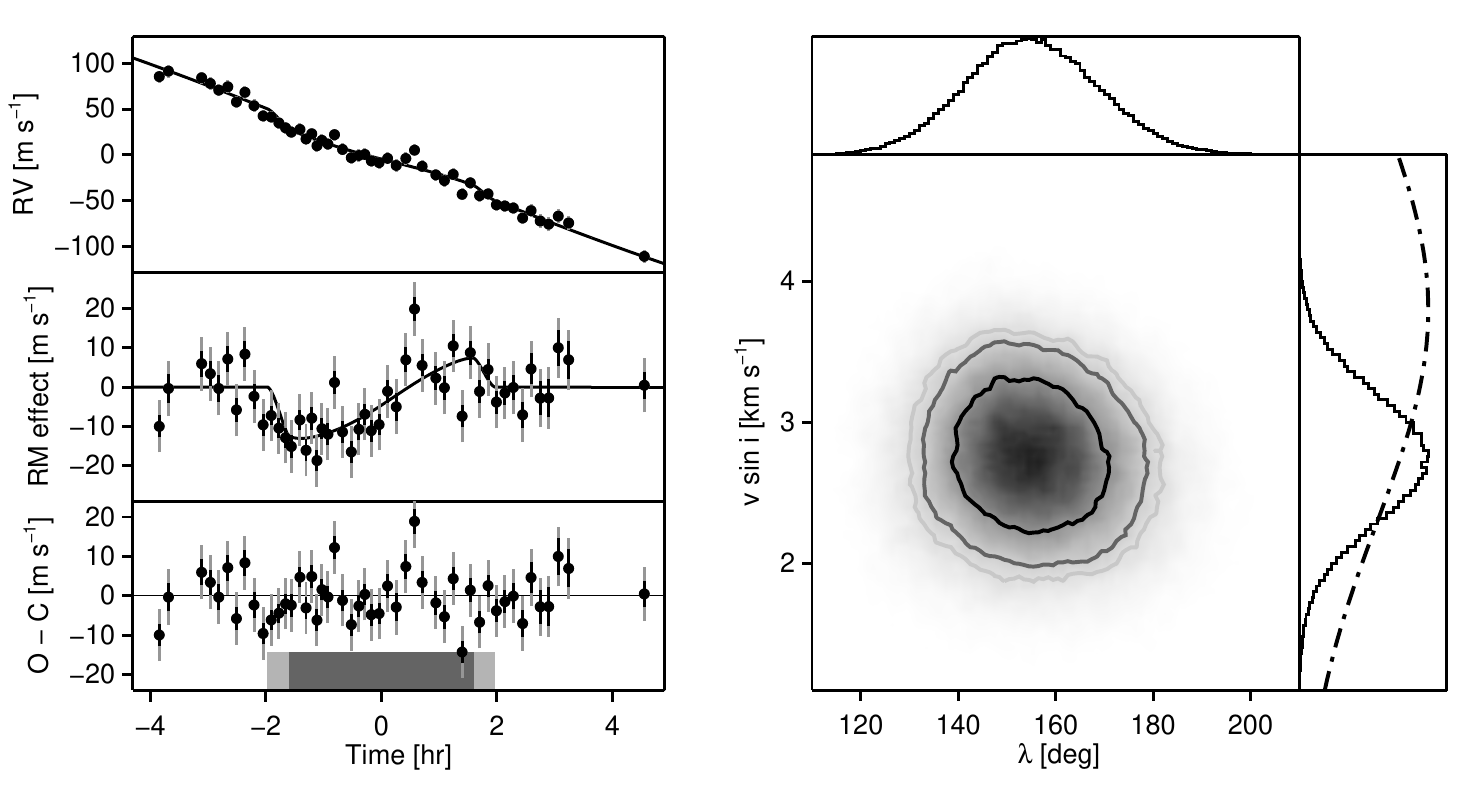}
  \caption {\label{fig:hatp7}  {\bf Spectroscopy of HAT-P-7 transit}
     Similar to Figure~\ref{fig:hatp6} but this time for data obtained
    during a transit in the HAT-P-7 system. In the middle panel on the
    left side, the orbital contribution to the observed RVs has been
    subtracted, isolating the RM effect.} 
  \end{center}
\end{figure*}

\section{Notes on individual systems}
\label{sec:systems}

In this section we report briefly on each of the 17 systems in our
sample (14 new observations and three re-analyses of previously
published data). We mention whenever we deviated from the general
approach described above. We also discuss briefly the values for $v
\sin i_{\star}$ and $\lambda$ that were obtained. For cases in which
$\lambda$ has been previously reported in the literature, we compare
the new value with the previous value.

\begin{figure*}
  \begin{center}
  \includegraphics{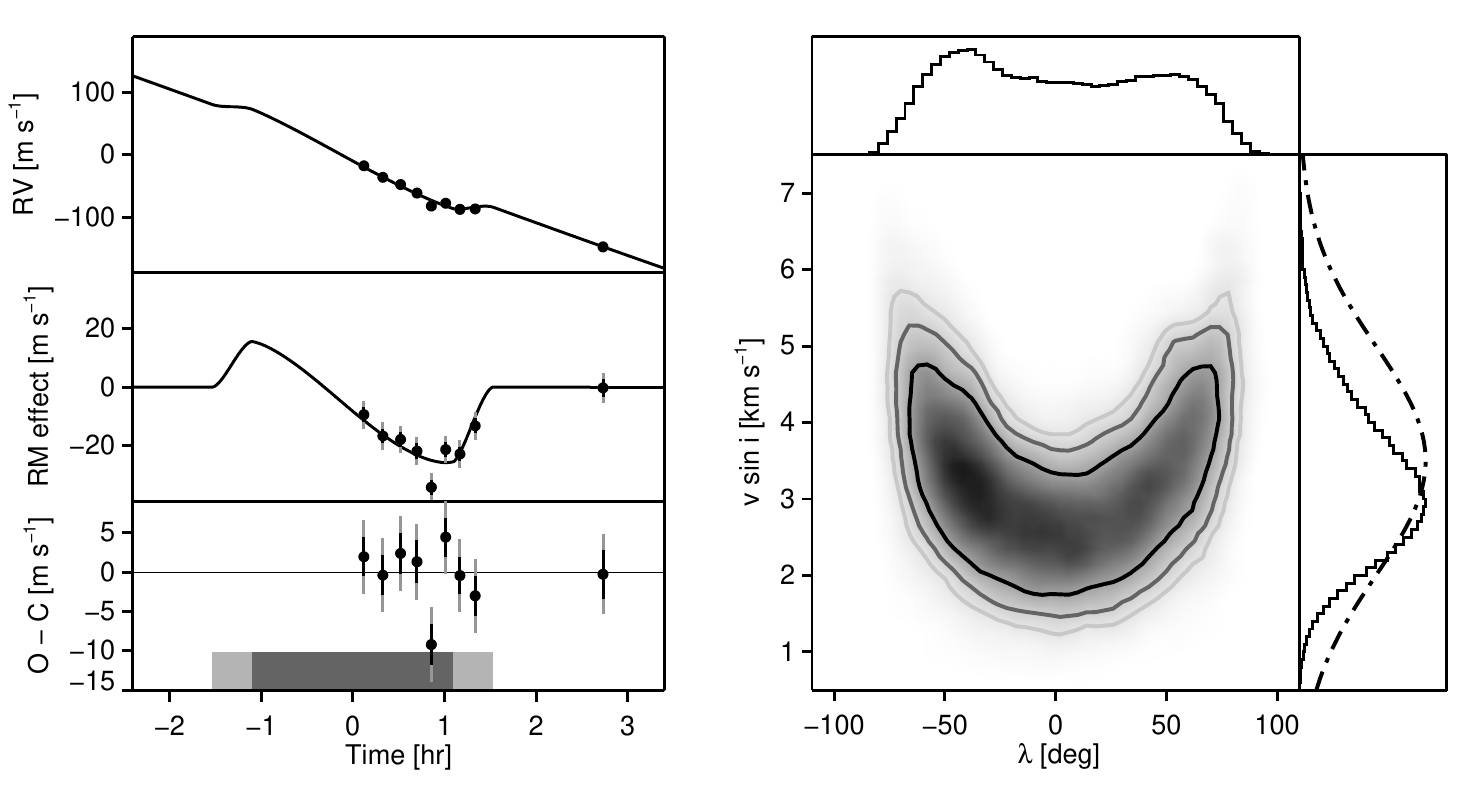}
  \caption {\label{fig:hatp16}  {\bf Spectroscopy of HAT-P-16 transit}
    Similar to Figure~\ref{fig:hatp7} but this time for data obtained
    during a transit in the HAT-P-16 system.} 
  \end{center}
\end{figure*}

\begin{figure*}
  \begin{center}
    \includegraphics{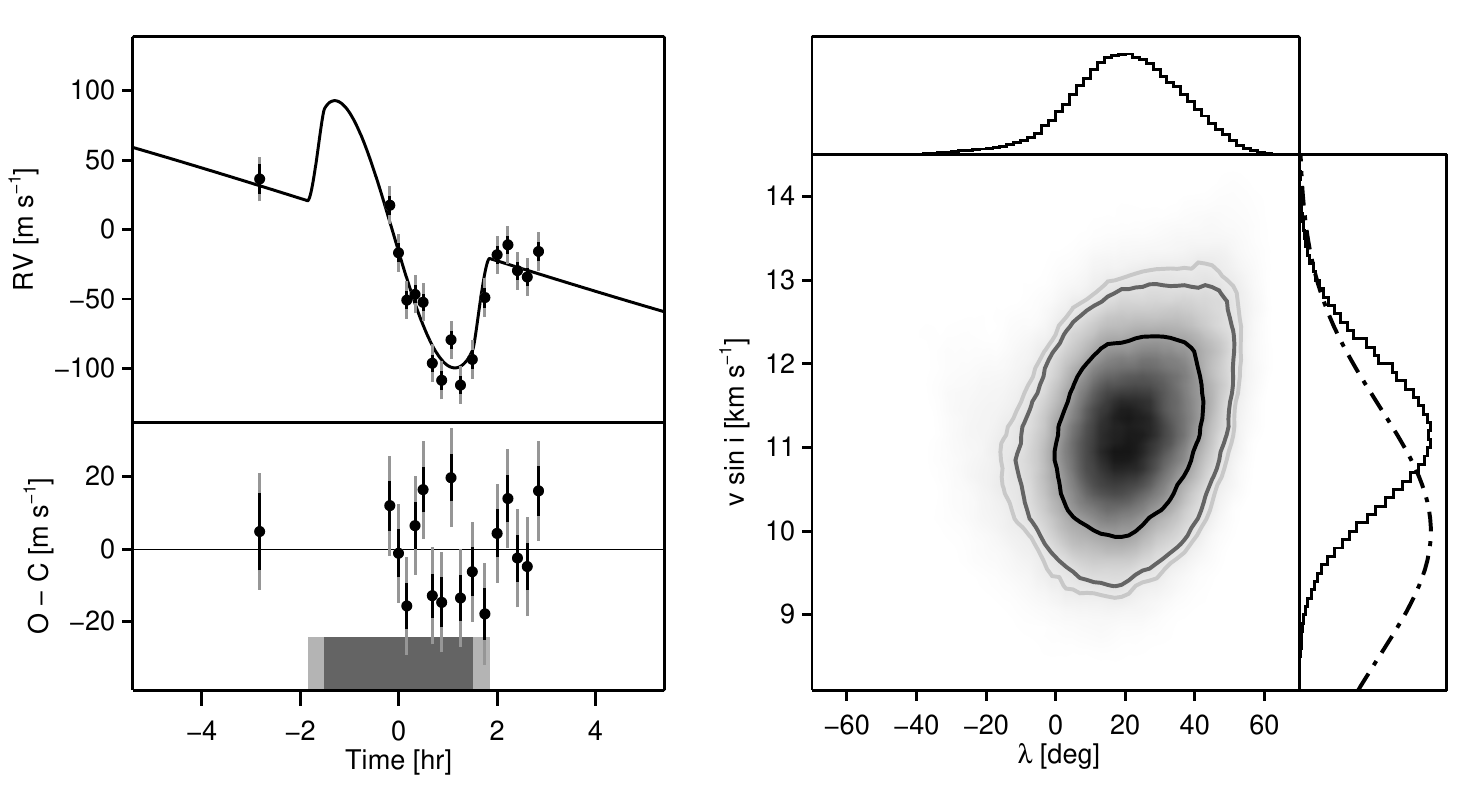}
    \caption {\label{fig:hatp24}  {\bf Spectroscopy of HAT-P-24 transit}
    Similar to Figure~\ref{fig:hatp6} but this time for data obtained
    during a transit in the HAT-P-24 system.} 
  \end{center}
\end{figure*}

\subsection{HAT-P-6}
\label{sec:hatp6}

The discovery of the hot Jupiter HAT-P-6\,b was reported by
\cite{noyes2008}.  We obtained 42 HIRES spectra over a five-hour
period covering the transit, which occurred on the night of
5/6~November~2010. Although little data could be obtained before
ingress, the observations continued for 1.5 hours after the 3.5-hour
transit. All the RVs are shown in the left panels of
Figure~\ref{fig:hatp6} along with the best-fitting model. The RV curve
immediately reveals a retrograde orbit.  The right panels show the
results in the $\lambda$ -- $v \sin i_{\star}$ plane. To obtain these
we used the HIRES RVs together with the prior information from
\cite{noyes2008} which are reported in Table~\ref{tab:priors}. We find
$\lambda=165\pm6^\circ$ and $v \sin i_{\star}=7.8\pm0.6$~km\,s$^{-1}$.
With this we confirm, with higher precision, the measurement presented
by \cite{hebrard2011b}. They found $\lambda = 166\pm10^\circ$ and $v
\sin i_{\star}=7.5\pm1.6$~km\,s$^{-1}$.

When the model was rerun with a prior constraint on $K_{\star}$, we
obtained $\lambda=175\pm4^\circ$, in disagreement with the prior-free
result and with the previously reported value.  (See
Table~\ref{tab:results} and Figure~\ref{fig:k1prior}.) Apparently the
star exhibits RV noise that is correlated on the timescale of at least
a few hours.

\begin{figure*}
  \begin{center}
  \includegraphics{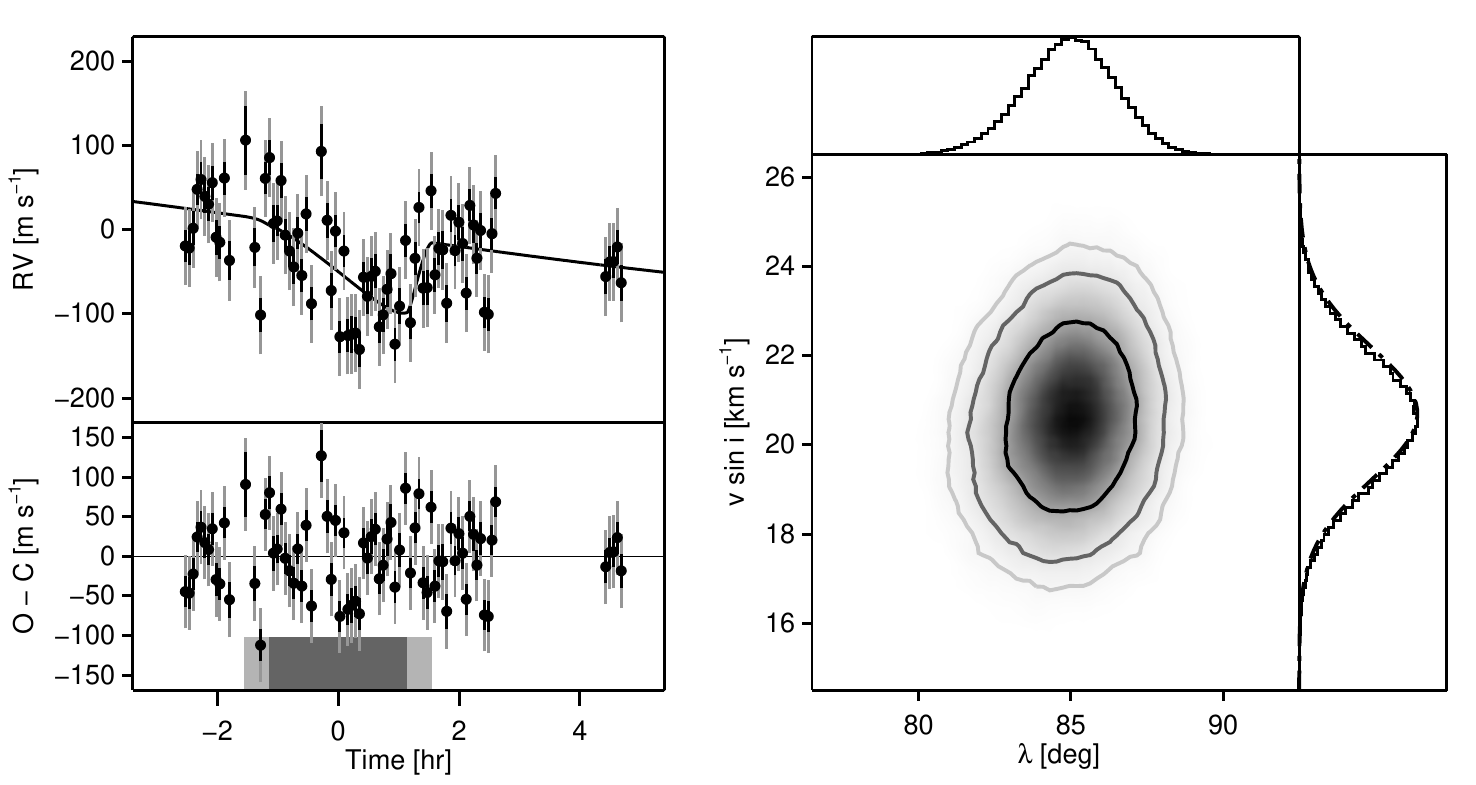}
  \caption {\label{fig:hatp32}  {\bf Spectroscopy of HAT-P-32 transit}
    Similar to Figure~\ref{fig:hatp6} but this time for data obtained
    during a transit in the HAT-P-32 system.} 
  \end{center}
\end{figure*}

\begin{figure*}
  \begin{center}
    \includegraphics{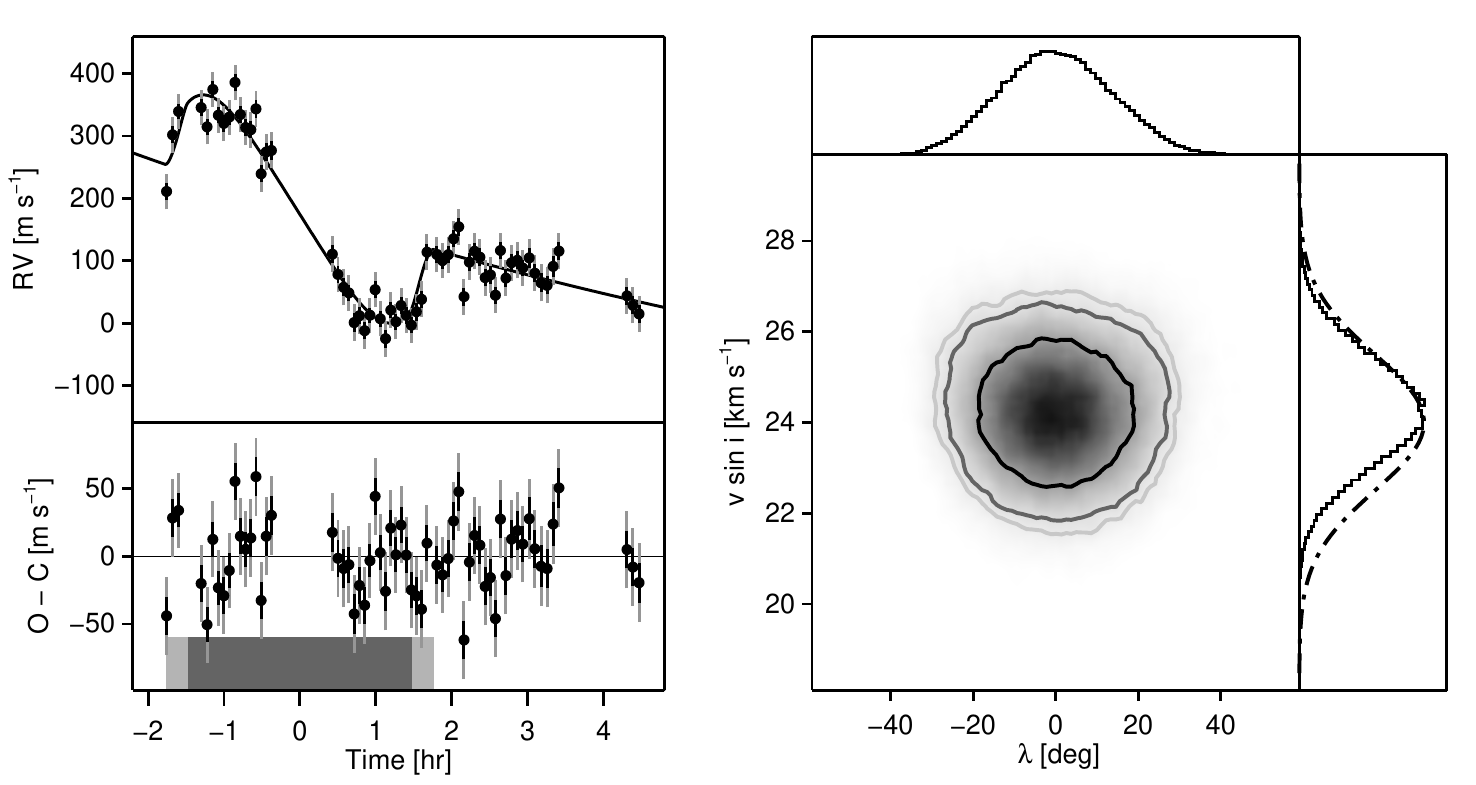}
    \caption {\label{fig:hatp34}  {\bf Spectroscopy of HAT-P-34 transit}
    Similar to Figure~\ref{fig:hatp6} but this time for data obtained
    during a transit in the HAT-P-34 system.} 
  \end{center}
\end{figure*}

\subsection{HAT-P-7}
\label{sec:hatp7}

\cite{pal2008} reported the discovery of HAT-P-7b. This was one of the
first systems for which a significant misalignment between the stellar
spin and orbital angular momentum was discovered in a star-planet
system \citep{winn2009,narita2009}. Both research groups reported an
apparently retrograde orbit. In fact this star is likely to be viewed
nearly pole-on, judging from its unusually low $v\sin i_\star$
\citep{winn2009,schlaufman2010}, but with nearly antiparallel {\it sky
  projections} of the orbital and stellar rotation vectors.

Interestingly the two independent results for $\lambda$ were in
disagreement, with \cite{winn2009} reporting $\lambda$ of $182.5
\pm9^\circ$ and $v \sin i_{\star}=4.9^{+1.2} _{-0.9}$~km\,s$^{-1}$,
and \cite{narita2009} reporting
$\lambda=-132.6^{+10.5}_{-16.3}$$^\circ$ (equivalent to
$\lambda=227.4^{+10.5}_{-16.3}$$^\circ$) and $v \sin
i_{\star}=2.3^{+0.6} _{-0.5}$~km\,s$^{-1}$. Both groups used the same
instrument (HDS on Subaru). With HIRES we reobserved the system
before, during, and after a planetary transit occurring during the
night 23/24~July~2010. The RVs are displayed in
Figure~\ref{fig:hatp7}. While the RM effect is clearly detected, the
SNR of the detection is low enough that an analysis of the RM effect
would benefit from a very precise prediction for the transit midpoint.
For this reason we downloaded the Quarter 6 Kepler light curve for
this system, covering the epoch of our spectroscopic transit
observations. The analysis of transit light curves is described in
section~\ref{sec:quantitative} and the derived photometric priors are
given in Table~\ref{tab:priors}. We obtain a $\lambda$ of
$155\pm14^\circ$ and a $v \sin i_{\star}=2.7\pm0.5$~km\,s$^{-1}$.

While the $v \sin i_{\star}$ is intermediate between the two values
reported previously, and the finding of a retrograde orbit is
confirmed, the value we obtained for $\lambda$ does not agree with
either of the previously reported values. What can have caused the
disagreement between the three results? While we can not give a
definite answer we suspect that two characteristics of this system
might have been important. $K_{\star}$ is ten times larger than the
amplitude of the RM effect making a clean separation of the RM effect
from the orbital radial velocity challenging. In addition the RM
amplitude is with $\sim20$~m\,s$^{-1}$ only a few times larger than
the typical uncertainty in the RV measurements. \cite{albrecht2011}
argued that for low SNR detections and for $\lambda$ near $0^{\circ}$
or $180^{\circ}$ the uncertainty in $\lambda$ can be
underestimated. However we cannot rule out the possibility that our
model is missing some aspect of stellar astrophysics, perhaps one that
is most apparent for nearly pole-on systems.  The HAT-P-7 system is
the only system in our sample where there exists a significant
disagreement between several independent measures of $\lambda$.

Since the discrepancies are unresolved, and to avoid
over-interpretation of the results in the subsequent discussion, the
uncertainty reported in Table~\ref{tab:results} is the standard
deviation ($37^\circ$) of the three independent measurements.

\begin{figure*}
  \begin{center}
  \includegraphics{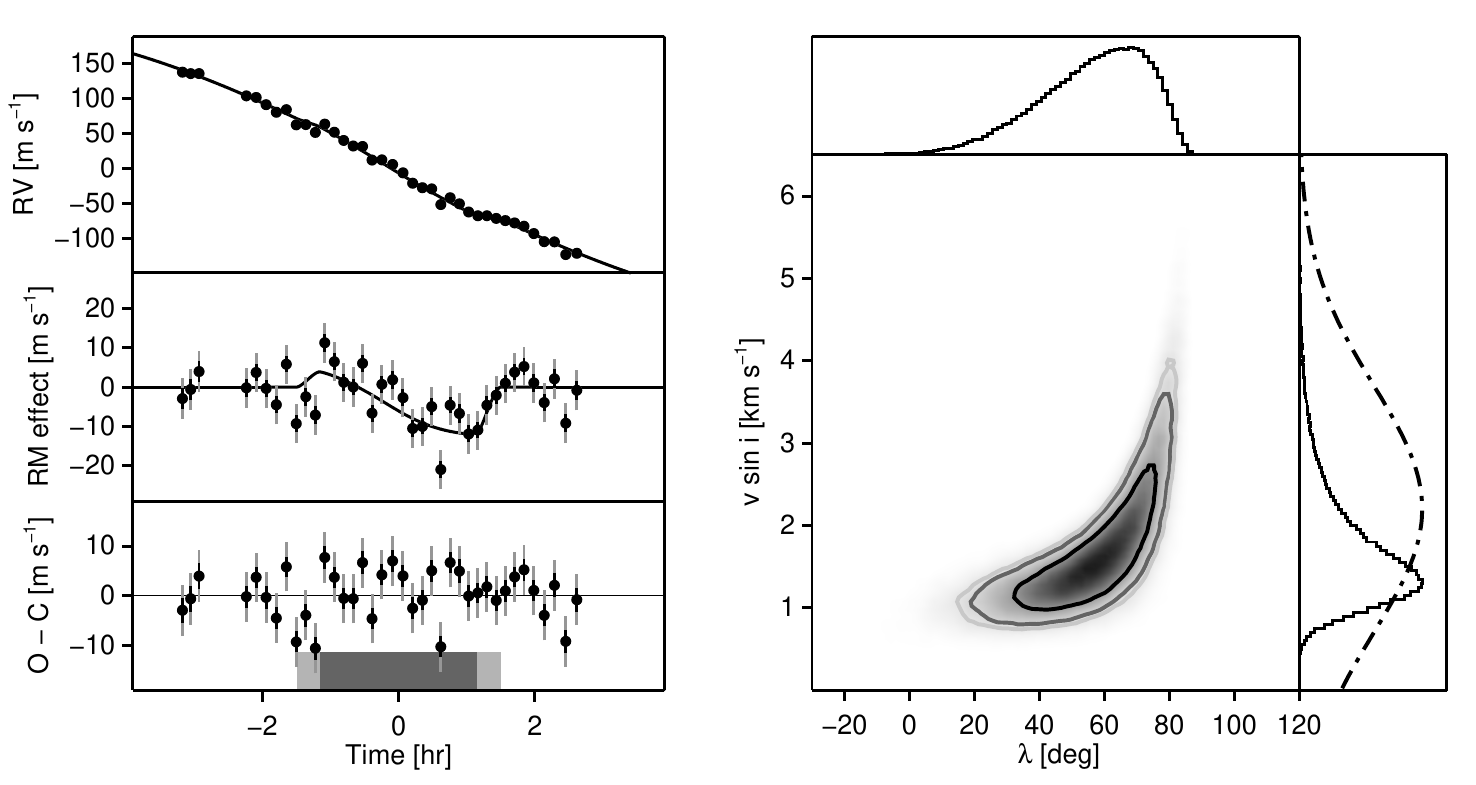}
  \caption {\label{fig:wasp12}  {\bf Spectroscopy of WASP-12 transit}
    Similar to Figure~\ref{fig:hatp7} but this time for data obtained
    during a transit in the WASP\,12 system.} 
  \end{center}
\end{figure*}

\begin{figure*}
  \begin{center}
  \includegraphics{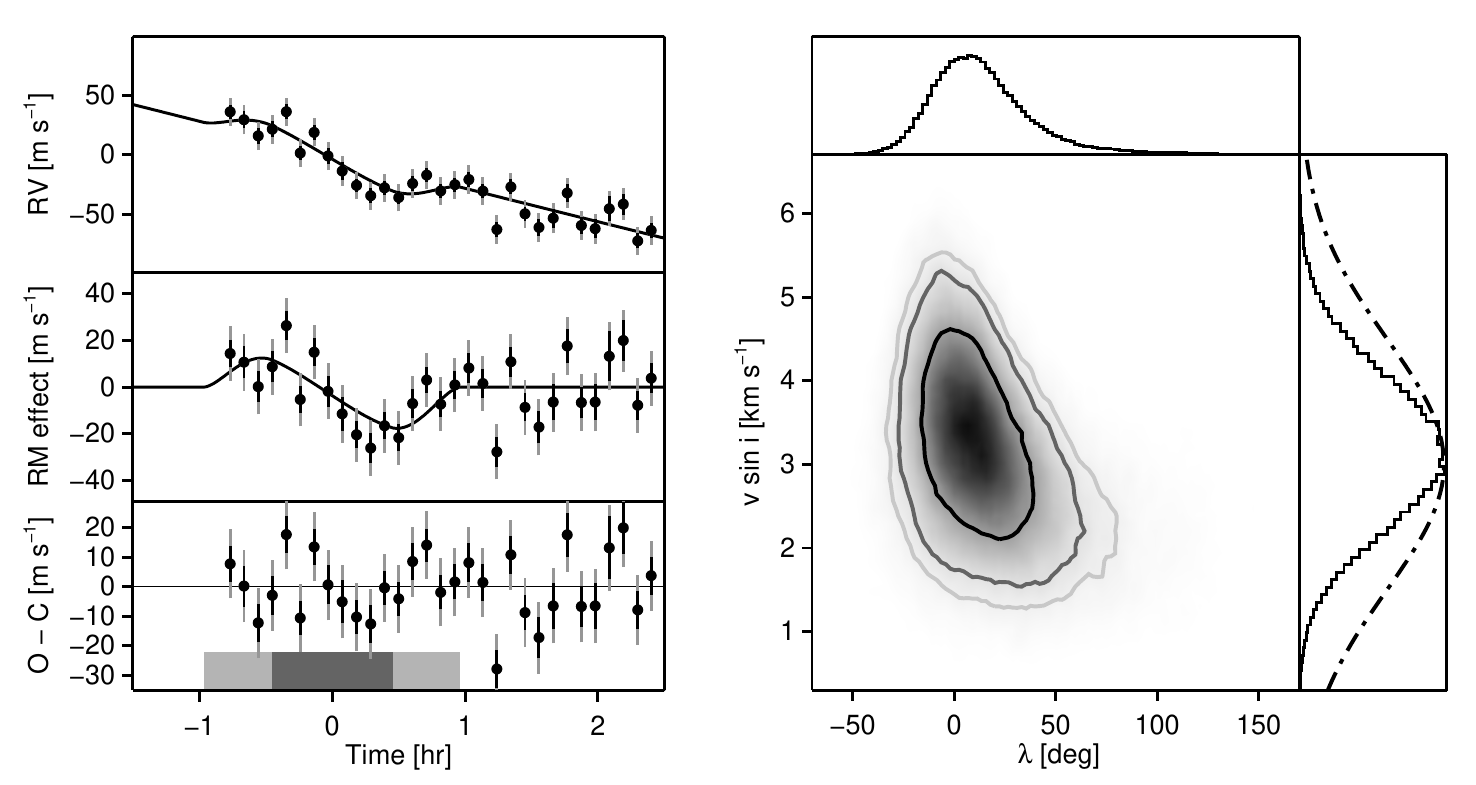} 
    \caption {\label{fig:wasp16}  {\bf Spectroscopy of WASP-16 transit}
    Similar to Figure~\ref{fig:hatp7} but this time for data obtained
    during a transit in the WASP\,16  system.} 
  \end{center}
\end{figure*}

\subsection{HAT-P-16}
\label{sec:hatp16}

The existence of HAT-P-16b was announced by \cite{buchhave2010}. We
observed this northern system with the Keck\,I telescope during the
night 26/27 December 2010. Due to bad weather we were only able to
observe the second half of the transit (Figure~\ref{fig:hatp16}). No data
before the transit were obtained. Using these RVs together with the
photometric and $v \sin i_{\star}$ ($3.5\pm1.5$~km\,s$^{-1}$) priors
from \cite{buchhave2010} we found that HAT-P-16\,b orbits its host
star on a prograde orbit ($\lambda=2^{+55}_{-46}$$^\circ$). While the
data are consistent with good alignment, little more can be
learned. Concerning $v \sin i_{\star}$ we found a value of
$3.1\pm1.0$~km\,s$^{-1}$, consistent with the prior. \cite{moutou2011}
found a projected obliquity of $\-10\pm16^\circ$ for this system. In
the subsequent discussion of Section~\ref{sec:discussion}, we adopt
their value.

\subsection{HAT-P-24}
\label{sec:hatp24}

The 24th HATNet planet was announced by \cite{kipping2010}. We
obtained HIRES spectra of this system during the night 27/28 September
2010. As was the case for HAT-P-16, the observations were interrupted
by bad weather, but at least for HAT-P-24 we obtained one pre-ingress
data point (see Figure~\ref{fig:hatp24}). This is why the result,
$\lambda = 20\pm16^\circ$, is much more precise than for
HAT-P-16. Nevertheless with the current data set is it not possible to
exclude a small misalignment.

\subsection{HAT-P-32}
\label{sec:hatp32}

HAT-P-32\,b was detected around a star which displays RV jitter on the
order of 80\,m\,s$^{-1}$ \citep{hartman2011}.  Therefore the SNR of
the RM effect detection in our dataset, obtained on 5/6 December 2011,
is relatively low (see Figure~\ref{fig:hatp32}). Nevertheless this is
the system in our sample for which we obtain the highest precision in
our measurement of the projected obliquity
($\lambda=85\pm1.5^\circ$). This curious fact is the consequence of
the system having a low impact parameter ($0.12$) {\it and} at the
same time presenting an asymmetric RM curve.

For systems with low impact parameters, even a strong misalignment
will lead mostly to a reduction of the amplitude of the RM effect,
with hardly any asymmetry in the RM curve.
\citep[e.g.][]{gaudi2007,albrecht2011b}. Only a narrow range of angles
near $90^\circ$ can produce a asymmetric RM effect. Therefore when an
asymmetry is detected in such a system, even relatively coarse RV data
will lead to high precision in $\lambda$.  Interestingly if we take
the $v \sin i_{\star}$ prior at face value and assume that our RM
model contains all the relevant physics, then we find with
$T_{21}=0.41251\pm0.0017$~hr a fivefold improvement in
our knowledge of the ingress duration compared to the prior constraint
($T_{21}=0.4128\pm0.0096$~hr). This is because for a given $v \sin
i_{\star}$ and $\lambda$ near $\pm 90^\circ$, the impact parameter is
encoded in the amplitude of the RM effect.

The pairing of HAT-P-32 and HAT-P-2 is instructive (see
Section~\ref{sec:hatp2}). For HAT-P-2 the RM effect was observed with
a very high SNR: the ratio between the amplitude of the RM effect and
the typical RV uncertainty in the RVs is $\sim10$. But despite the
very high SNR in the detection of the RM effect, we obtained a
relatively low precision in the measurement of $\lambda$.  This is the
reverse of what we see for HAT-P-32.  For this system, the ratio
between the amplitude of the RM effect and the average RV uncertainty
is $\sim2$.  Nevertheless we obtain a very high precision in the
measurement of $\lambda$. This highlights how important the geometry
of the transit {\it together} with the projected obliquity is in
determining the pression in the measurement of the projected
obliquity.

\begin{figure*}
  \begin{center}
   \includegraphics{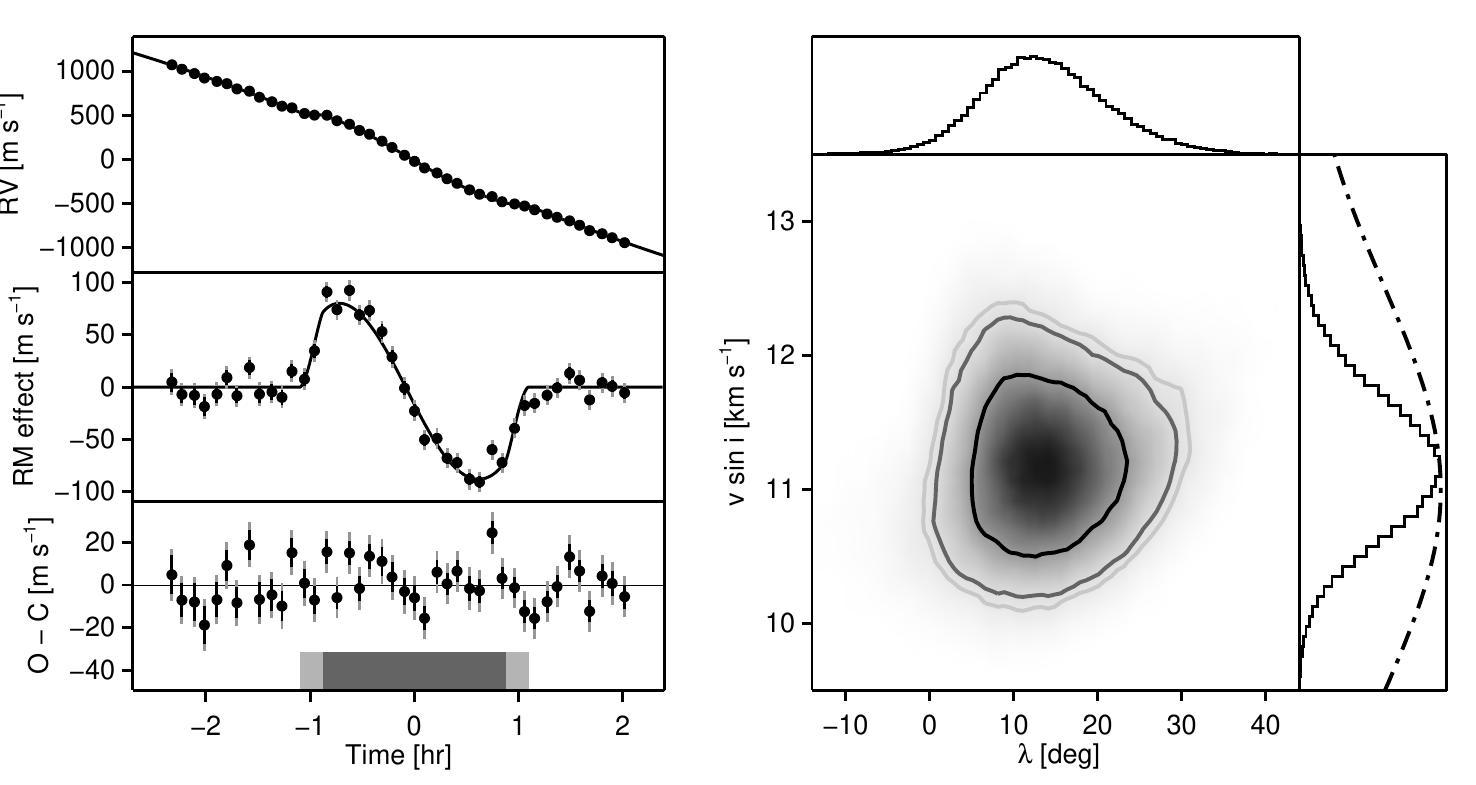}
  \caption {\label{fig:wasp18}  {\bf Spectroscopy of WASP-18 transit}
    Similar to Figure~\ref{fig:hatp7} but this time for data obtained
    during a transit in the WASP\,18 system.} 
  \end{center}
\end{figure*}

\subsection{HAT-P-34}
\label{sec:hatp34}

HAT-P-34\,b was announced by \cite{bakos2012} together with three
other transiting planets. HAT-P-34\,b orbits its host star on a highly
eccentric ($e=0.44$) orbit with a relatively long period of
$5.5$~days.  We observed this system with HIRES during the night 2/3
September 2011.  We obtained 62 spectra during and after the transit
(Figure~\ref{fig:hatp34}). As the RM curve suggests we find that the
projections of the stellar rotation and orbital angular momenta are
well aligned ($\lambda=0\pm14^\circ$).

\begin{figure*}
  \begin{center}
  \includegraphics{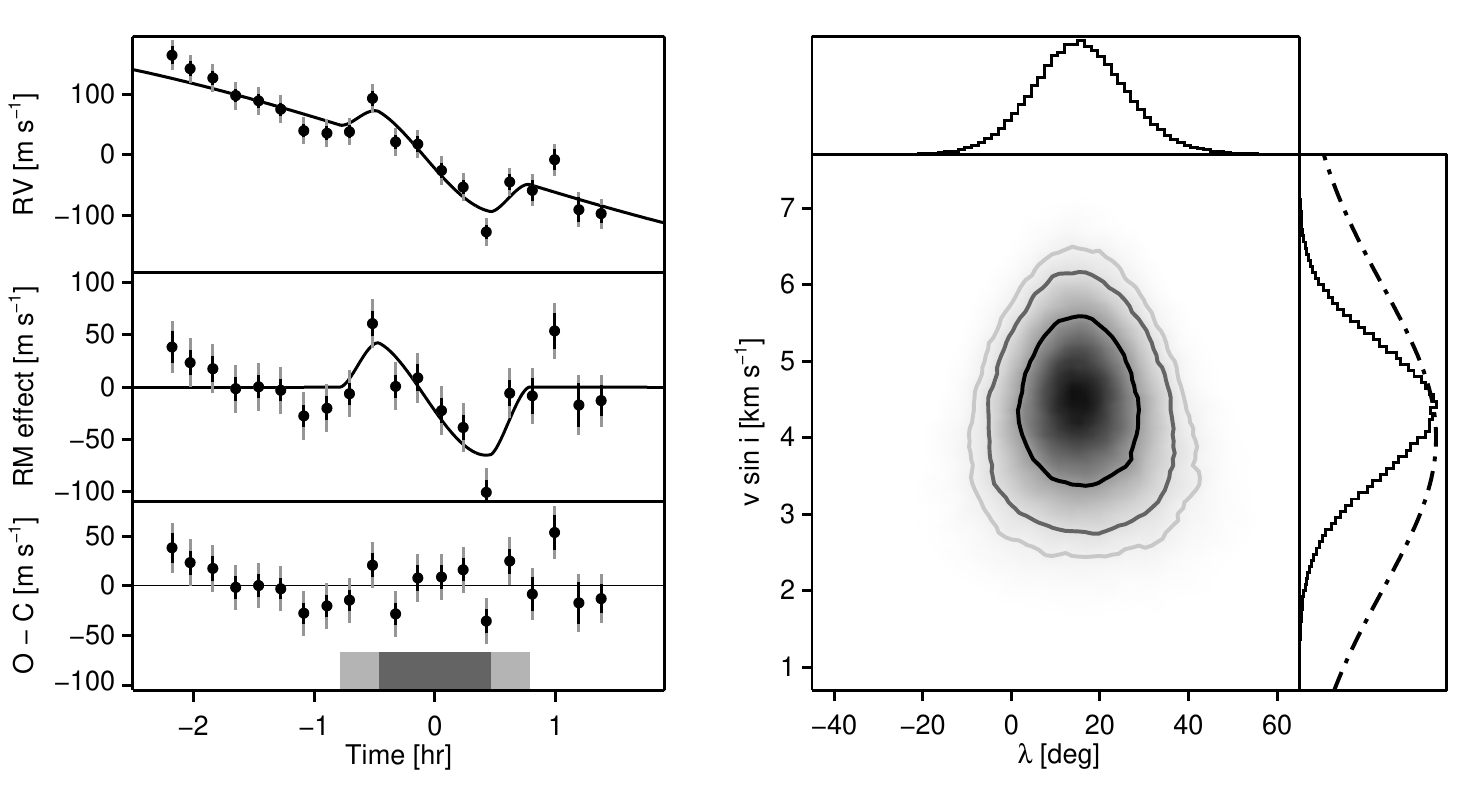}
     \caption {\label{fig:wasp19}  {\bf Spectroscopy of WASP-19 transit}
    Similar to Figure~\ref{fig:hatp7} but this time for data obtained
    during a transit in the WASP\,19 system.} 
  \end{center}
\end{figure*}

\begin{figure*}
  \begin{center}
   \includegraphics{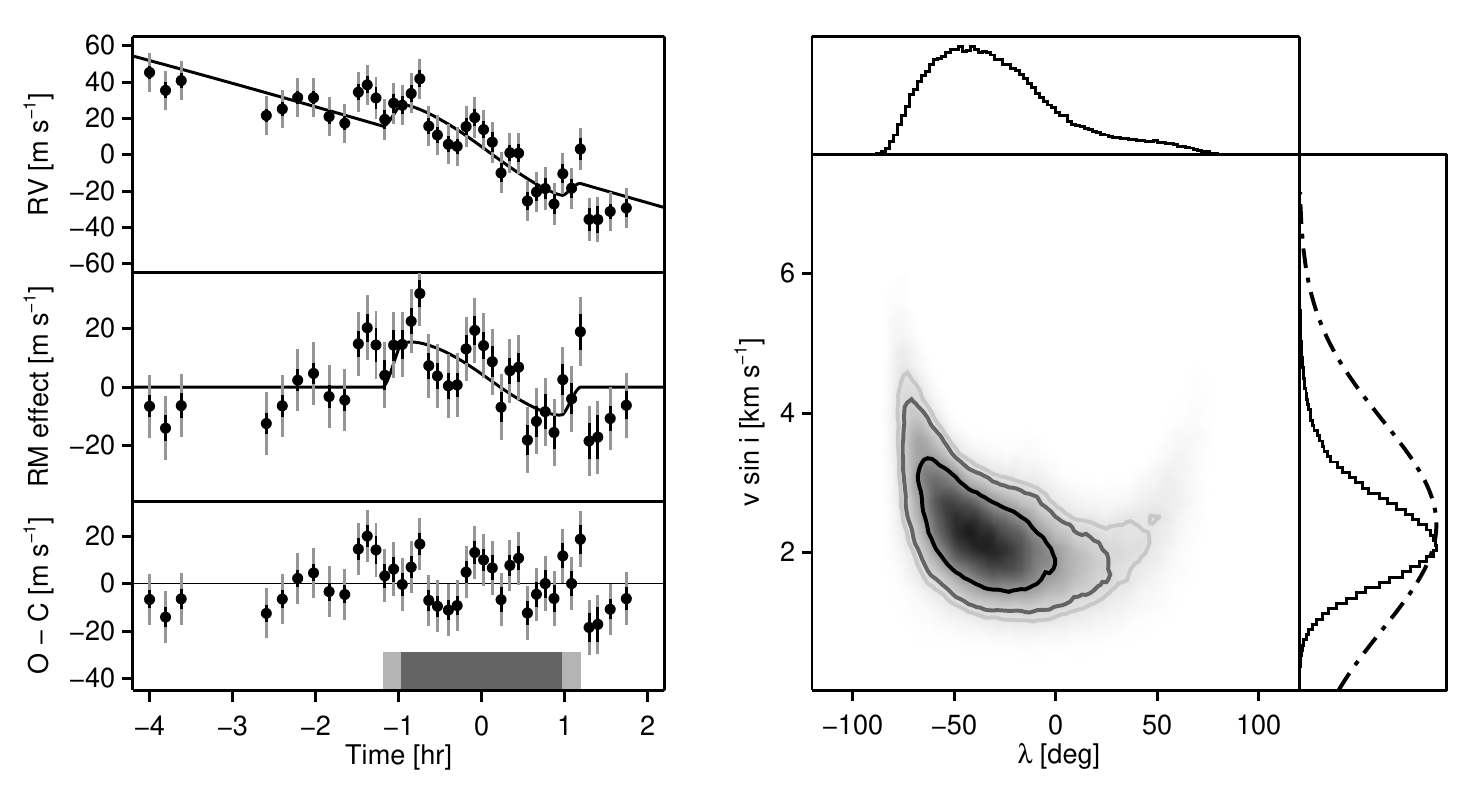} 
  \caption {\label{fig:wasp26}  {\bf Spectroscopy of WASP-26 transit}
    Similar to Figure~\ref{fig:hatp7} but this time for data obtained
    during a transit in the WASP\,26 system.} 
  \end{center}
\end{figure*}

\subsection{WASP-12}
\label{sec:wasp12}

WASP-12 harbors a transiting exoplanet with a short orbital period of
$1\fd1$. It was discovered by \cite{hebb2009}. We took from those
authors our prior on $v \sin i_{\star}$, while we obtained from
\cite{maciejewski2011} our priors on the geometric transit
parameters.\footnote{\cite{Husnoo2011} also obtained RV data during a
  transit of this system, but their results were not conclusive. }

We observed the RM effect in this system on 1/2 January 2012 with the
Keck\,I telescope. Our analysis of the 38 RVs
(Figure~\ref{fig:wasp12}, left panel) obtained before, during, and
after the 3~hr transit indicates a misalignment of
$59^{+15}_{-20}$$^\circ$ and $v \sin
i_{\star}=1.6^{+0.8}_{-0.4}$~km\,s$^{-1}$. For this system there is a
strong correlation between $\lambda$ and $v \sin i_{\star}$; for
higher projected rotation speeds $\lambda$ increases and approaches
$90^\circ$ (Figure~\ref{fig:wasp12}, right panel). This is interesting
as WASP-12, with a mass of $1.35$~$M_{\odot}$ \citep{maciejewski2011},
is expected to be a fast rotator and to have a high $v$.  Using the
method employed by \cite{schlaufman2010} we find an expected $v$ of
$13.7\pm2.5$~km\,s$^{-1}$ suggesting a low $\sin i_{\star}$, i.e., the
stellar spin axis is inclined along the line of sight.  It seems very
likely that the star and the planetary orbit are misaligned, though we
cannot tell with certainty how much of the misalignment is in the
plane of the sky as opposed to the perpendicular direction. A similar
situation was found for WASP-1 \citep{albrecht2011b}.

\subsection{WASP-16}
\label{sec:wasp16}

\cite{lister2009} found this transiting hot Jupiter on a 3.1-day orbit
around a southern-sky star. We used the PFS in conjunction with the
Magellan\,II telescope to obtain RVs during a transit occurring 3/4
June 2010 (Figure~\ref{fig:wasp16}).  We used the ephemeris and
information on the projected rotation speed from the discovery paper
as prior information in the fit. As no information on $T_{21}$ was
given, we used the EULERCAM light curve presented in the discovery
paper to establish priors for $T_{41}$, $T_{21}$, and $R_{\rm
  p}/R_{\star}$. This system is one of those for which we find (based
on the fit to the out-of-transit observations) an orbital velocity
semi-amplitude ($K_{1}=353\pm54$~m\,s$^{-1}$) that is significantly
different from the orbital solution presented in the discovery paper
($K_{1}=116.7^{+2.4}_{-1.9}$~m\,s$^{-1}$). To investigate this issue
we obtained 10 out-of-transit observations on a number of different
nights. These observations agree with the previously reported orbital
solution. A large stellar spot might be responsible for the excess RV
change during the transit night (see section~\ref{sec:rv_var}),
although such a spot would need to cover a substantial part of the
stellar surface.

\begin{figure*}
  \begin{center}
   \includegraphics{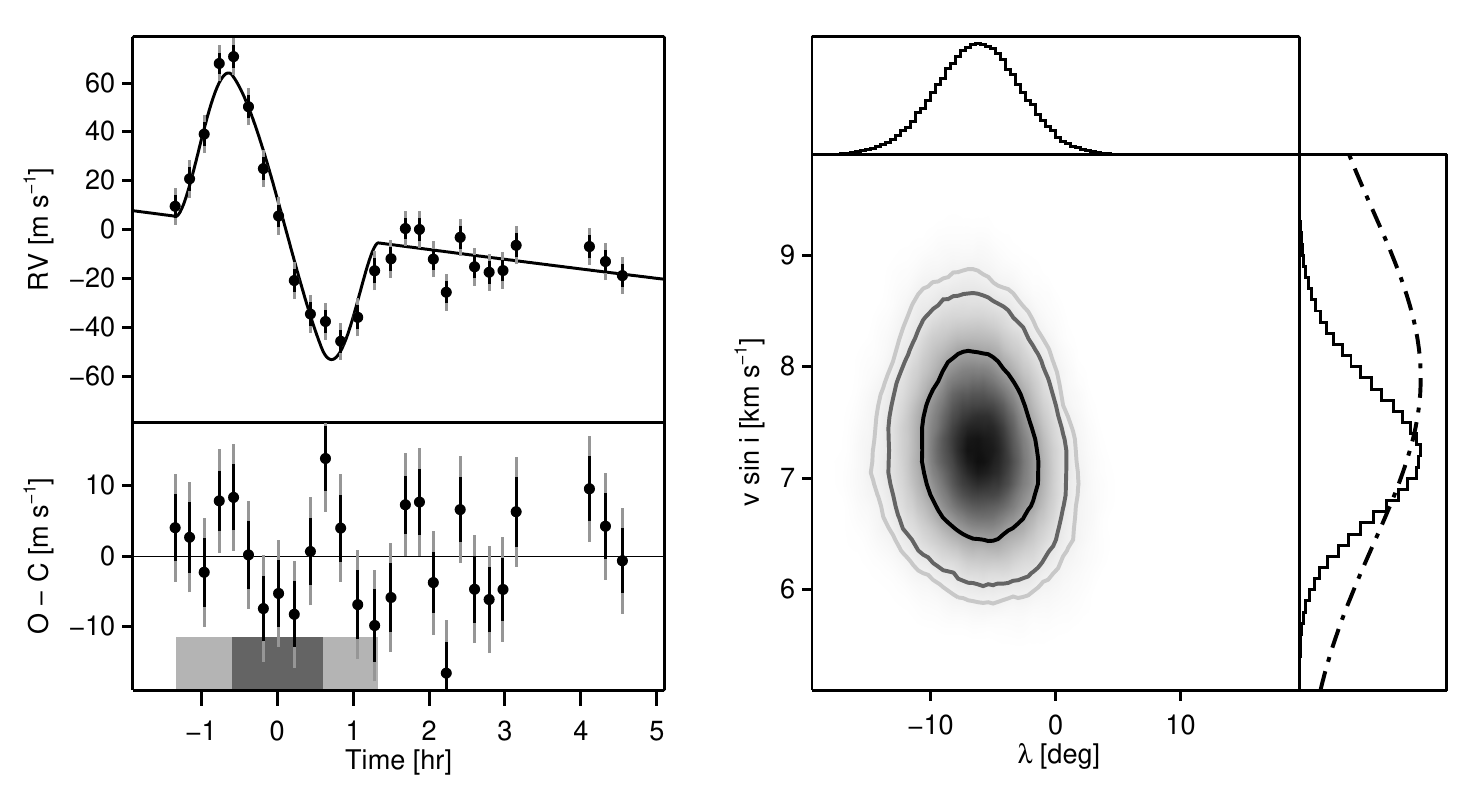} 
  \caption {\label{fig:wasp31}  {\bf Spectroscopy of WASP-31 transit}
    Similar to Figure~\ref{fig:hatp6} but this time for data obtained
    during a transit in the WASP\,31 system.} 
  \end{center}
\end{figure*}

We omitted 9 data points obtained at the end of the transit night,
after the template observation for WASP-16 was obtained. These
observations would in principle facilitate a separation of the RM
signal from other sources of RV variation. However, given that there
is already an indication of intrinsic RV variations apart from orbital
motion, and the absence of any guarantee that the intrinsic variations
are linear in time over the course of many hours, we decided to use
only the data points immediately following egress.  All the RVs are
listed in Table~\ref{tab:rvs}.  Using the 31 remaining RV data points
obtained during and after the transit we obtained $\lambda=
11^{+26}_{-19}$$^\circ$ and $v \sin
i_{\star}=3.2\pm0.9$~km\,s$^{-1}$. \cite{brown2012} recently measured
$\lambda= -4.2^{+11}_{-13.9}$$^\circ$ consistent with the value
obtained here. However they found a significantly lower value for $v
\sin i_{\star}$ ($1.2^{+0.4}_{-0.5}$~km\,s$^{-1}$) which is not
consistent with their spectroscopic prior, nor with ours. We also note
that \cite{brown2012} did not report any disagreement between the
out-of-transit velocity gradient observed on the transit night, and
the published orbital solution.

\begin{figure*}
  \begin{center}
  \includegraphics{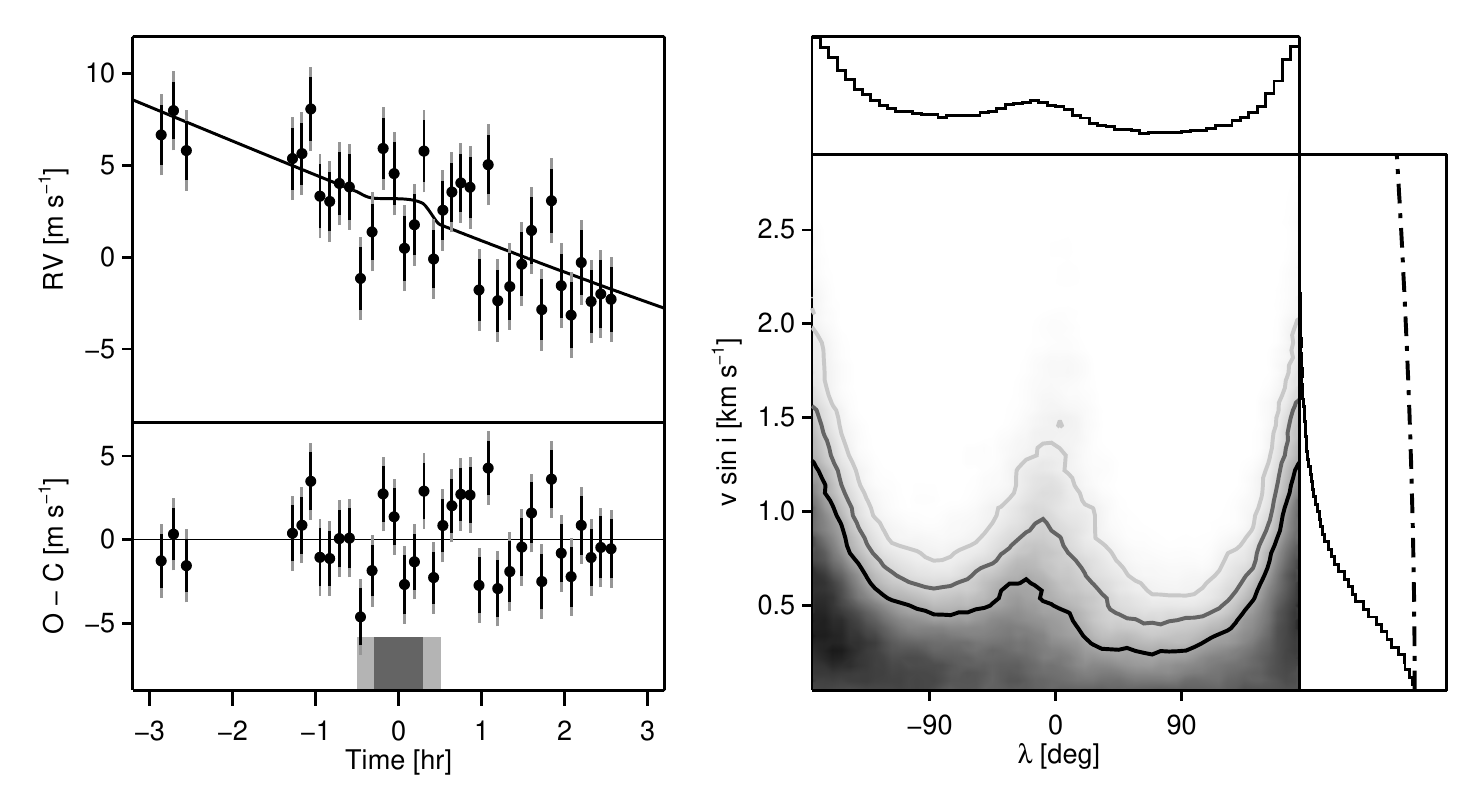}
  \caption {\label{fig:gi436} {\bf Spectroscopy of Gl\,436 transit}
    Similar to Figure~\ref{fig:hatp6} but this time for data obtained
    during a transit in the GI\,436 system. }
  \end{center}
\end{figure*}

\begin{figure*}
  \begin{center}
  \includegraphics{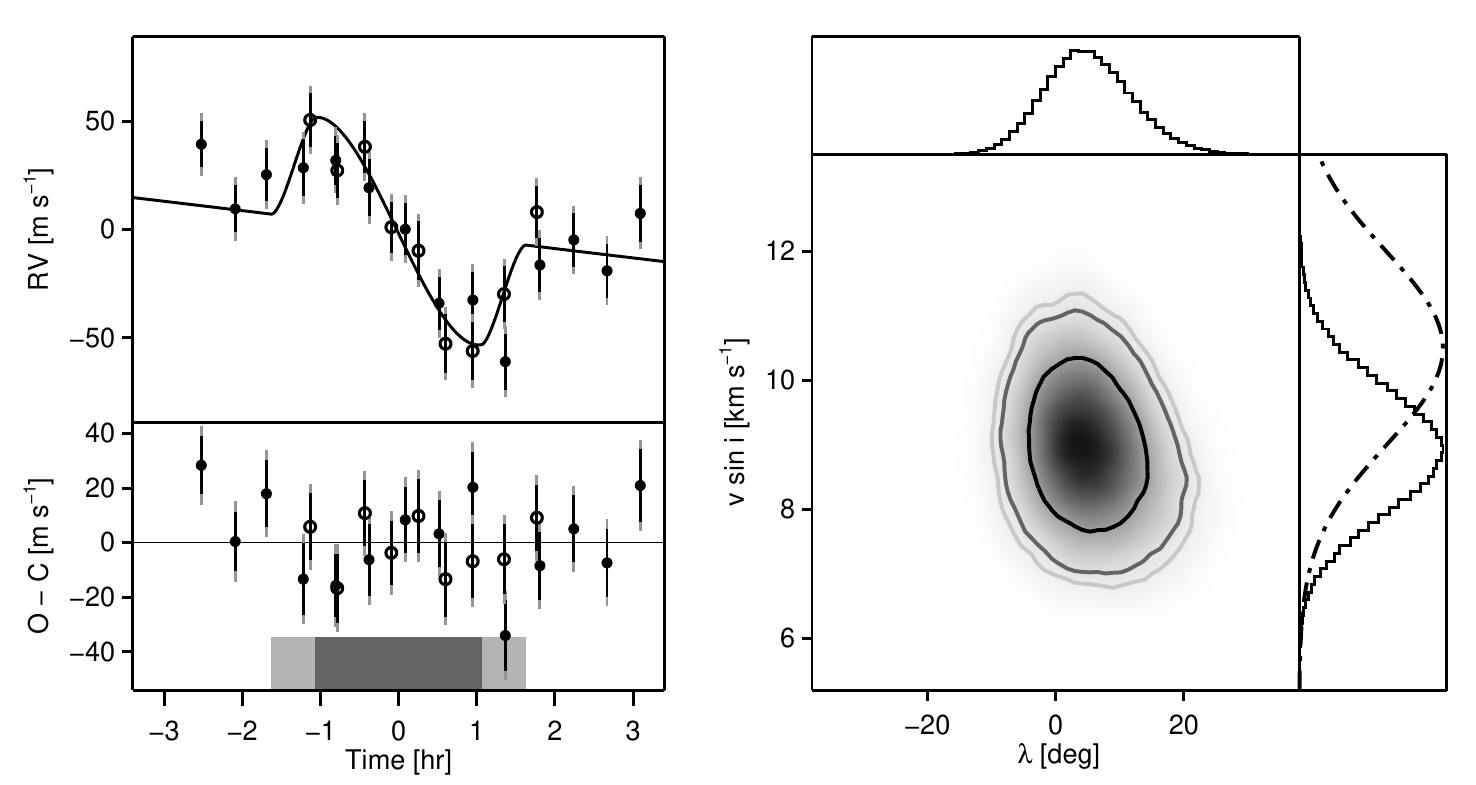}
  \caption {\label{fig:kepler8}  {\bf Spectroscopy of Kepler-8 transit}
    Similar to Figure~\ref{fig:hatp6} but this time for data obtained
    during a transit in the Kepler\,8 system. Black filled circles represent
    the RV data points already obtained by \cite{jenkins2010}, while the
    open circles show the new HIRES data.} 
  \end{center}
\end{figure*}

\begin{figure*}
  \begin{center}
 \includegraphics{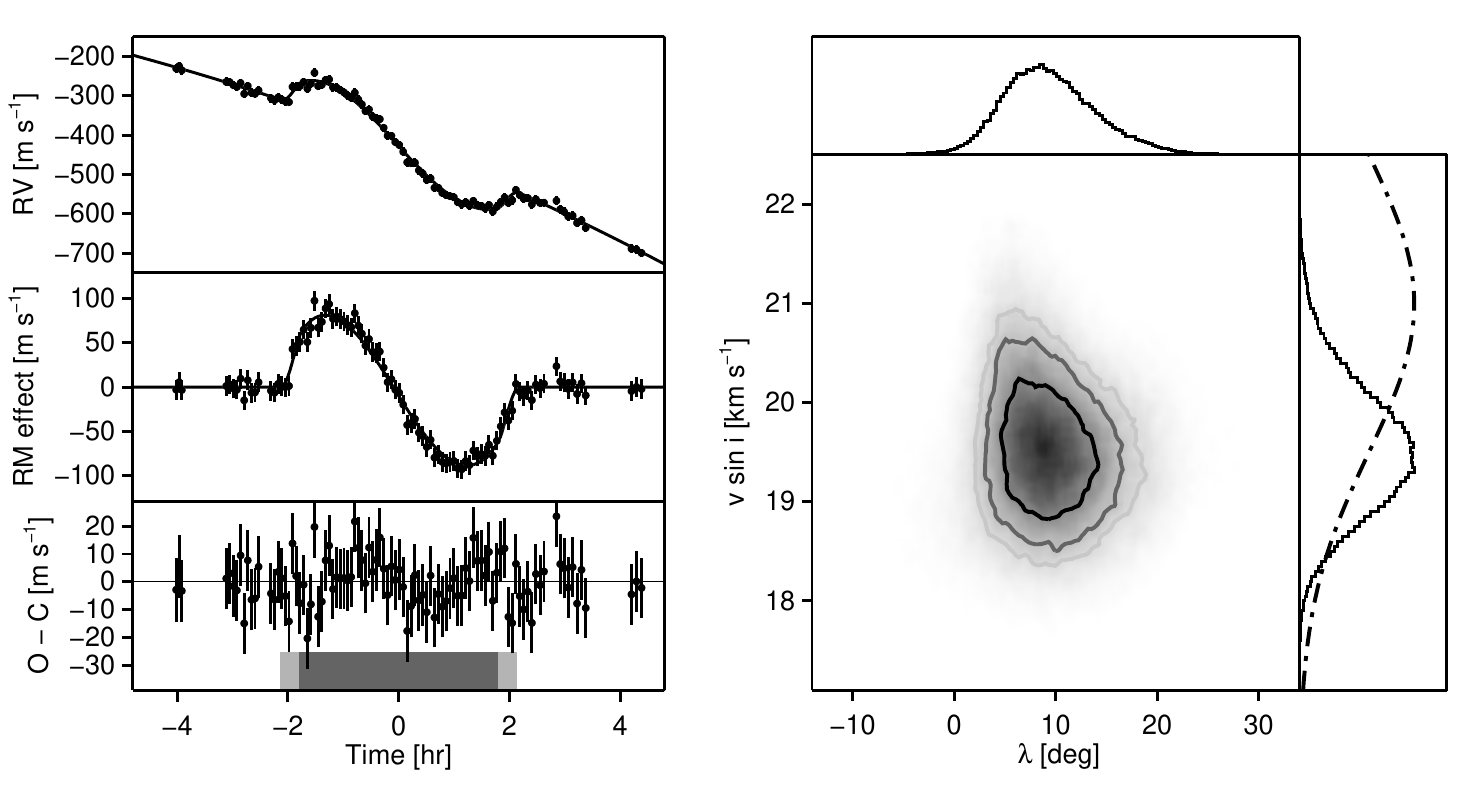}
  \caption {\label{fig:hd147506} {\bf Spectroscopy of HAT-P-2 transit}
   Similar to Figure~\ref{fig:hatp7} but this time for data obtained
    during a transit in the HAT-P-2 system. Structure can be seen in the
    residuals indicating that our modeling does not capture all the
    relevant physics of the RM effect. }
  \end{center}
\end{figure*}

\begin{figure*}
  \begin{center}
  \includegraphics{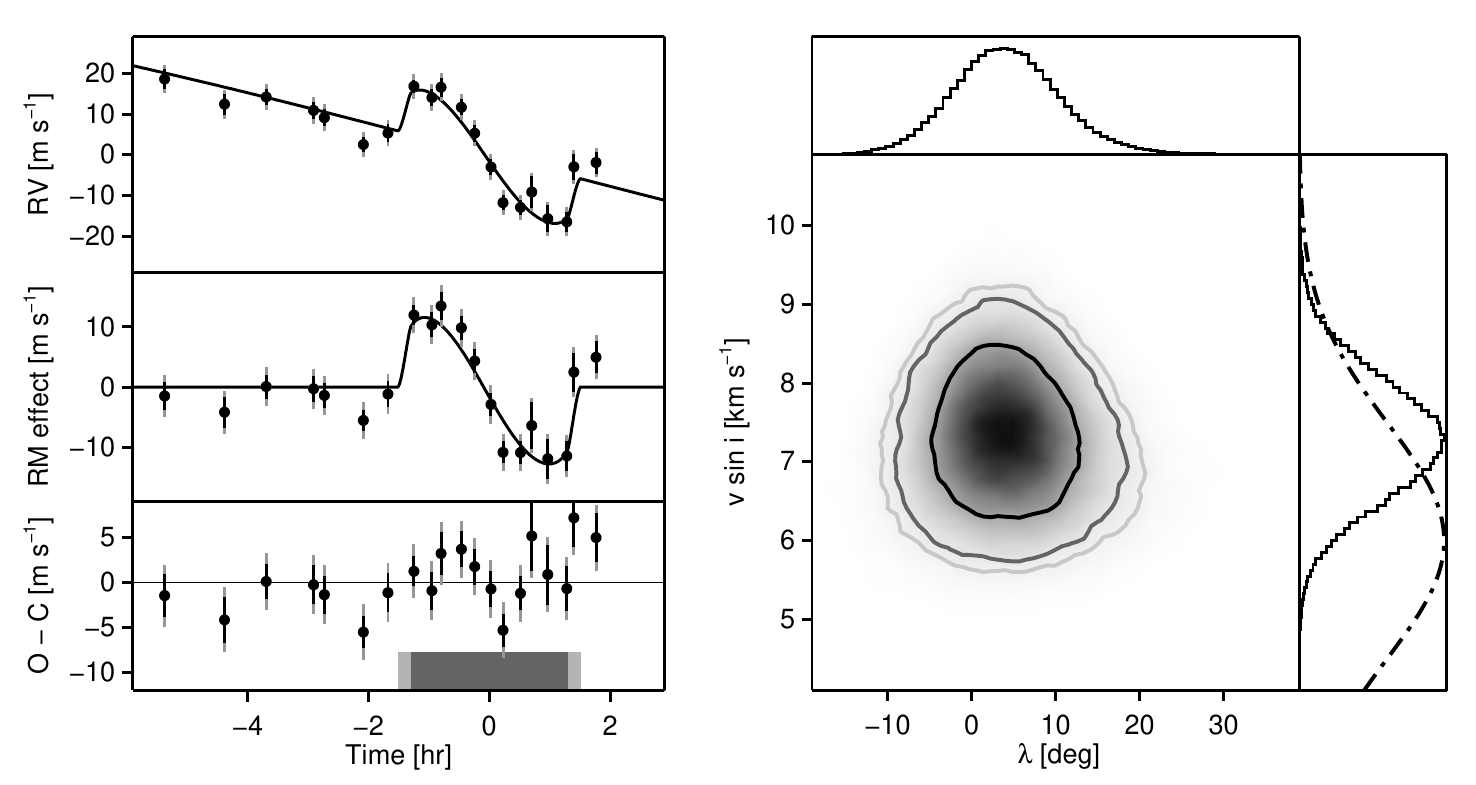}
  \caption {\label{fig:hd149026}  {\bf Spectroscopy of HD\,149026 transit}
    Similar to Figure~\ref{fig:hatp7} but this time for data obtained
    during a transit in the HD\,149026 system.} 
  \end{center}
\end{figure*}

\begin{figure*}
  \begin{center}
  \includegraphics{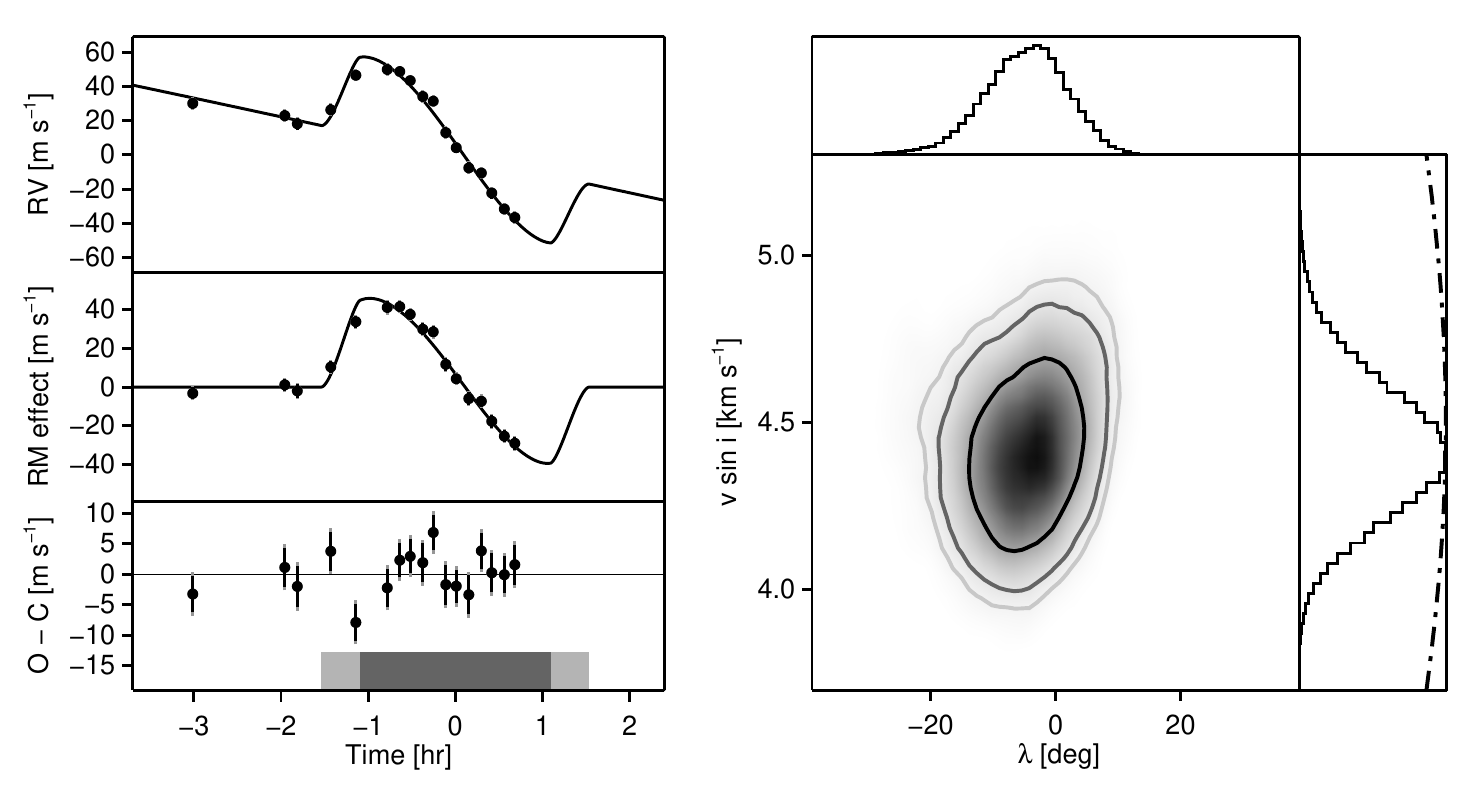}
  \caption {\label{fig:hd209458}  {\bf Spectroscopy of HD\,209458  transit}
    Similar to Figure~\ref{fig:hatp7} but this time for data obtained
    during a transit in the HD\,209458 system.} 
  \end{center}
\end{figure*}

\subsection{WASP-18}
\label{sec:wasp18}

WASP-18\,b orbits its host star in only $0.94$~days
\citep{hellier2009}. As noted by the discoverers it is an excellent
system to study the tidal interaction between a gas giant and its host
star. \cite{triaud2010} found the projections of the stellar spin and
orbital axes to be aligned ($\lambda=4\pm5^{\circ}$).

We obtained 48 data points with the PFS during the night 8/9 October
2011.  Six of these data points were obtained at a significant
distance in orbital phase from the transit, and are not included in the
fit, for reasons discussed in the preceding section and
section~\ref{sec:rv_var}. Figure~\ref{fig:wasp18} presents the
remaining 42 RVs obtained before, during, and after the transit.

No value for $T_{21}$ has been reported in the literature for
WASP-18\,b. We obtained the photometry from \cite{southworth2009} and
performed our own analysis. For the reasons mentioned by
\cite{southworth2010c} we did not use this light curve to improve on
the ephemeris for this system. The photometric priors reported in
Table~\ref{tab:priors} for this system are derived from the fit to
that light curve.
 
We found that the projected stellar spin ($\lambda=13\pm7^\circ$)
appears to be misaligned with the orbital rotation axis, although the
statistical significance is marginal. The value for $v \sin i_{\star}$
$11.0\pm0.5$~km\,s$^{-1}$ does not give any strong indication of an
inclination along the line of sight. Any obliquity in WASP-18 seems to
be small.

\subsection{WASP-19}
\label{sec:wasp19}

WASP-19 is another system with a very short-period ($0.78$~d) hot
Jupiter discovered by the WASP consortium \citep{hebb2010b}. We
observed this system with PFS during the transit night 20/21 May 2010
(Figure~\ref{fig:wasp19}).  Our photometric priors were based on
results by \cite{hellier2011}. These authors derived the projected
obliquity in the same fit that was used to determine the photometric
parameters. This could in principle make our observation of $\lambda$
dependent on their obliquity observation. However given the quality of
their photometric and RV data, we expect that the RVs obtained by them
have little influence on their results for $T_{41}$, $T_{21}$, and
$R_{\rm p}/R_{\star}$. We therefore expect our result
($\lambda=15\pm11^\circ$) to be independent of their result
($\lambda=4.6\pm5.2^\circ$). Our prior on the projected rotation speed was
taken from the discovery paper.

As can be seen in Figure~\ref{fig:wasp19} there is correlated excess
noise in the RVs during pre-egress. This together with the low SNR
detection raises the concern that we might not have detected a RM
effect at all, and we are fitting our RM model to noise. We carried
out an experiment similar to the one described by \cite{albrecht2011b}
for the WASP-2 system.  Briefly, we create $200\,000$ fake datasets
with similar noise characteristics as the real dataset, but without an
underlying RM effect. We then fitted an RM model to each dataset. Form
the resulting density distribution we conclude that our RM detection
represents a 3$\sigma$ detection. We also used the Bayesian
Information Criterion (BIC) to assess whether the model including the
RM effect is preferred over a model without the RM effect
\citep[e.g.][]{brown2012}. The RM model is preferred, with BIC$_{\rm
  RM}=48$ and BIC$_{\rm no\,RM}=60$. However the BIC test depends
somewhat on how many of the RV measurements were obtained during
transit, compared to the number obtained outside of
transit. Furthermore both tests (or at least our implementations of
these tests) assume that the noise is uncorrelated, which does not
seem to be the case for the present dataset.

\subsection{WASP-26}
\label{sec:wasp26}

This system was detected by \cite{smalley2010}. \cite{anderson2011}
attempted a measurement of the RM effect but did not make a secure
detection. We obtained 39 RV data points during the night 17/18 August
2010 (Figure~\ref{fig:wasp26}). Using these RVs, photometric priors
from \cite{anderson2011}, and a prior on the projected stellar
rotation speed from \cite{smalley2010}\footnote{ We used the $v \sin
  i_{\star}$ value from \cite{smalley2010} as prior, and not the value
  reported by \cite{anderson2011}. This is because the former authors
  estimated $v\sin i_\star$ under the assumption that the
  macroturbulent parameters is $4.1\pm 0.3$~km\,s$^{-1}$. This is
  closer to our estimation ($5.1\pm1.5$~km\,s$^{-1}$) than the value
  of $3.0\pm 0.3$~km\,s$^{-1}$ used by the latter authors.} we derive 
$\lambda=-34^{+36}_{-26}$$^\circ$ and $v \sin
i_{\star}=2.2\pm0.7$~km\,s$^{-1}$.

\subsection{WASP-31}
\label{sec:wasp31}

The existence of WASP-31\,b was announced by \cite{anderson2011c}. We
observed this system with the Keck\,I telescope during the night 12/13
March 2012 and obtained 27 RV measurements (Figure~\ref{fig:wasp31}). 

Using these RVs together with the $v \sin i_{\star}$, $T_{41}$,
$T_{21}$, and $R_{\rm p}/R_{\star}$ priors from \cite{anderson2011c}
and priors on the ephemeris from \cite{dragomir2011} we found a low
projected obliquity ($\lambda=-6\pm3^\circ$). We obtained $v \sin
i_{\star}=7.3\pm0.4$~km\,s$^{-1}$, consistent with the prior
($7.9\pm1.5$~km\,s$^{-1}$). \cite{brown2012} recently found
$\lambda=2.8\pm3.1^\circ$ and $v \sin
i_{\star}=7.5\pm0.7$~km\,s$^{-1}$. The value for $\lambda$ is
inconsistent with our result. \cite{brown2012} used the ephemeris
presented in \cite{anderson2011c}. Using the $T_c$ and $P$ values from
\cite{anderson2011c} to set the constraint on the time of inferior
conjunction for the observed spectroscopic transit we obtain
$\lambda=2\pm3^\circ$. This is consistent with the result by
\cite{brown2012}. For our final results we decided to use the timing
information from \cite{dragomir2011}, who used the ephemeris from the
discovery paper in combination with an additional light curve
(although we note that the new light curve contains only a small
amount of post-egress data).

Therefore while we have excellent agreement between two different
research groups, the dependence of the result on the photometric
priors and the inconsistency between the results using the same RVs
but different timing information cast some doubt on the formal
uncertainty intervals. For this reason, we double the uncertainty in
$\lambda$ before including this system in the subsequent discussion of
the interpretation of all the results. Future photometric observations
will be helpful, but at least it seems clear that the projected
obliquity is small.

\subsection{Gl\,436}
\label{sec:gi436}

The transiting planet in Gl\,436 was discovered by \cite{butler2004}
and found to be transiting by \cite{gillon2007}. It would be of
interest to know the obliquity in this system as Gl\,436\,b is of
similar mass then HAT-P-11\,b, which is so far the only Neptune-class
planet for which the host star's obliquity has been measured. Using
HIRES we obtained RVs of the system during the night of
24/25~April~2010. To reduce the uncertainties in the photometric
parameters, we gathered new photometric data with Keplercam, a CCD
camera on the $1.2$~m telescope of the Fred L.\ Whipple Observatory on
Mount Hopkins, Arizona \citep{szentgyorgyi2005}. The RV measurements
are displayed in the left panels of Figure~\ref{fig:gi436}. The right
panels display the posterior in the $\lambda$ -- $v \sin i_{\star}$
plane. No RM effect was detected, and therefore the data provide no
constraint on $\lambda$. Only an upper limit of $0.4$~km\,s$^{-1}$ on
$v \sin i_{\star}$ can be obtained. This upper limit on $v \sin
i_{\star}$ is more constraining than the prior we adopted, which was a
one-sided Gaussian prior enforcing $v \sin i_{\star} <
5$~km\,s$^{-1}$. Thus from the transit data we learned only that the
star is a slow rotator. The structure in the posterior distribution is
caused by the lower sensitivity of our measurement for nearly prograde
and retrograde configurations. See \cite{albrecht2011b} for an
explanation of this effect.

\subsection{Kepler-8}
\label{sec:kepler8}

\cite{jenkins2010} reported the discovery of Kepler-8 and also found
$\lambda=-26\pm10.1^{\circ}$, a misalignment between the projections
of the stellar and orbital spins. As the data set available to
\cite{jenkins2010} only consisted of data taken during the transit
itself, it is difficult to assign a significance to the result. We
therefore obtained a new data set with HIRES during the transit night
7/8 August 2011. We used both HIRES data sets in our analysis,
including two additional parameters representing a second velocity
offset and a second RV transit-night velocity gradient between the two
different transit nights. Flat priors were assigned to these
parameters.  We found that the data are consistent with good alignment
between the stellar and orbital rotation axes (see
Figure~\ref{fig:kepler8}). We obtained $\lambda=5\pm7^\circ$ and $v
\sin i_{\star}= 8.9\pm1.0$~km\,s$^{-1}$. The velocity offset between
the two data sets was found to be $12\pm10$~m\,s$^{-1}$. Using only
the new dataset in our analysis, we obtained
$\lambda=5\pm10^\circ$. Using only the older dataset, we obtained
$\lambda=30^{+55}_{-28}$$^{\circ}$. We chose to use both datasets for
this system, as they have been obtained with the same instrument and
setup, and the RVs were extracted with the same code, circumstances
different from all the other systems for which multiple transits have
been observed. Furthermore this case highlights the importance of
prior constraints on the data. Without a constraint on the RV offset
as applied by \cite{jenkins2010}, the original dataset is
consistent with good alignment.

\subsection{HAT-P-2}
\label{sec:hatp2}

HAT-P-2 was one of the first systems for which a measurement of the
projected obliquity was reported. \cite{winn2007} measured $\lambda=
1.2\pm13.4^{\circ}$. \cite{loeillet2008} found $\lambda=
0.2^{+12.2}_{-12.5}$$^{\circ}$, a similar value. The uncertainty in
this measurement is relatively high because HAT-P-2\,b has a low
impact parameter. After these results were published, new and improved
system parameters based on new photometry were reported by
\cite{pal2010}. For this reason we reanalyzed the RVs obtained by
\cite{winn2007} together with the new photometric priors, obtaining
$\lambda=9\pm5^{\circ}$. Thanks to the improved photometric
information, the formal uncertainty in $\lambda$ is now lower, and
supports a 1.8$\sigma$ ``detection'' of a misalignment, a different
conclusion from the one by \cite{winn2007}. However, one can see in
the lower left panel of Figure~\ref{fig:hd147506} that, after
subtraction of our best fitting model, structure remains in the
residuals.

This is a clear sign that our model does not capture all the effects
influencing the RV anomaly during transit. We therefore consider the
uncertainty of our analysis to be itself quite uncertain, by at least
a factor of two. In Table~\ref{tab:results} we report a doubled
uncertainty relative to our formal results for $\lambda$ and $v \sin
i_{\star}$. Therefore while the photometric information as well as the
quality of the RV data would allow for a measurement of $\lambda$ to
within a few degrees, the limitations of our model of the RM effect
prevent this gain from being fully realized.  The model expects a
larger RM effect shortly before and after mid-transit and a lower
amplitude around the second and third contacts.  One possible effect
that would produce this type of structured residuals is a change of
limb darkening with depths of stellar absorption lines.  We note that
no such structure in the residuals was seen in the original analysis
of these data using a different modeling for the RM effect, which was
derived empirically for the specific spectrograph and stellar type
\citep{winn2007}.

It is not surprising that the limitations of our model would be most
apparent for HAT-P-2, as it has the combination of very rapid rotation
and very high signal-to-noise ratio, given the brightness of the star
($V=8.7$). For this system it would be better to analyze the
absorption line profiles directly, and their changes over the course
of the night, rather than deriving and modeling RVs
\cite[e.g.][]{albrecht2007,cameron2010}.

\subsection{HD\,149026}
\label{sec:hd149026}

\cite{wolf2007} found $\lambda =-12\pm15^\circ$ for this well-known
system. Since that time, \citep{carter2009} presented a more precise
light curve based on {\it Hubble Space Telescope} (HST) infrared
observations. Thus a re-analysis is warranted. We used the photometric
information from \cite{carter2009} and the $v \sin i_\star$ prior from
the discovery paper \citep{sato2005}. We found a projected obliquity
of $12\pm7^\circ$. This is only marginally consistent with the
previously reported value of $\lambda$ using the same RV data. The
main reasons for the difference are the use of new photometric data,
and the lack of post-egress data (Figure~\ref{fig:hd149026}) in
combination with our choice to not impose a prior constraint on
$K_{\star}$. When we repeated the analysis with a prior on $K_{\star}$
(Table~\ref{tab:results}) we found the results to be consistent with
those of \cite{wolf2007}.

\subsection{HD\,209458}
\label{sec:hd209458}

The first system in which a transiting planet was discovered was also
the first system for which the projected obliquity was measured.
\cite{queloz2000} found an angle consistent with alignment to within
$20^{\circ}$. \cite{winn2005} found $\lambda$ to be
$-4.4\pm1.4^\circ$. The later result suggested that a small but
significant misalignment exists in this system, similar to the
misalignment of our Sun against the ecliptic
($\sim7^{\circ}$). However, \cite{winn2005} used data from different
nights, unlike our current procedure. If the star is active or has
star spots, the long-term intrinsic RV noise would be manifested as
velocity offsets between different nights, which were not taken into
account by \cite{winn2005}. Indeed we find evidence that the stellar
jitter has different amplitudes on different timescales: by fitting a
model to all of the available HIRES data, we find the rms residual of
the data obtained on different nights to be $5$~m~s$^{-1}$ while the
rms residual of the transit-night data is only $3$~m~s$^{-1}$.

We therefore repeated the analysis using only the HIRES RVs obtained
during the transit night, 29/30 July 2000. We also used the HST light
curve obtained by \cite{brown2001} to obtain photometric priors, and
used the $v \sin i_\star$ value from \cite{laughlin2005} as a
prior. We obtained a projected stellar obliquity of $-5\pm7^\circ$.
The looser bounds on $\lambda$ are a consequence of the lack of egress
and post-egress data on the night 29/30 July 2000
(Figure~\ref{fig:hd209458}). This case emphasizes the need to obtain
data outside of the transit, unless one is willing to assume that the
long-term RV noise is negligible.

\begin{table}
  \begin{center}
    \caption{Measured projected obliquities \label{tab:proj_obli}}
    \smallskip 
    \begin{tabular}{l c
        r@{$\pm$}l  r@{$\pm$}l 
        r  }
      \tableline\tableline
      \noalign{\smallskip}
         & System 
         &  \multicolumn{2}{c}{  $v \sin i_{\star}$} & \multicolumn{2}{c}{ $\lambda$} 
         & ref \\
         &
         & \multicolumn{2}{c}{(km\,s$^{-1}$)} &\multicolumn{2}{c}{($^{\circ}$)}  
         & \\
      \noalign{\smallskip}
      \hline
      \noalign{\smallskip}
      1 & CoRoT-2 b &    10.9&0.5 & \multicolumn{2}{c}{$4.0^{+5.9}_{-6.1}$}  & 1 \\
      2 & CoRoT-3 b &    17.0&1.0 & \multicolumn{2}{c}{$37.6^{+10}_{-22.3}$} & 2,3 \\
      3 & CoRoT-18 b &    8.0& 1.0 &           -10& 20 & 4 \\
      4 & HAT-P-1 b &      3.8& 0.6 &           3.7&  2.1 &  5 \\
      5 & HAT-P-2 b &    19.5& 1.4 &              9&  10 & this work \\
      6 & HAT-P-4 b &       5.8& 0.3 &        -4.9& 11.9 &  6 \\
      7 & HAT-P-6 b &        7.8& 0.6 &         165&  6 &  this work\\
      8 & HAT-P-7 b &        2.7& 0.5 &          155& 37 & this work \\
      9 & HAT-P-8 b &      14.5& 5 &        \multicolumn{2}{c}{$-17^{+9.2}_{-11.5}$}  &  7 \\
      10 & HAT-P-9 b &    12.5& 1.8 &      -16&  8 & 7  \\
      11 & HAT-P-11 b &     1.0& 0.9 &    \multicolumn{2}{c}{$103^{+26}_{-10}$}  & 8 \\
      12 & HAT-P-13 b &     1.7& 0.4 &     1.9&  8.6 &  9 \\
      13 & HAT-P-14 b &     8.2& 0.5 &  -170.9&  5.1 & 6 \\
      14 & HAT-P-16 b &     3.9& 0.8 &   -10& 16 & 7  \\
      15 & HAT-P-23 b &     7.8& 1.6 &    15& 22 & 7 \\
      16 & HAT-P-24 b &    11.2& 0.9 &    20& 16 & this work \\
      17 & HAT-P-30 b &     3.1& 0.2 &    73.5&  9.0 &  10 \\
      18 & HAT-P-32 b &   20.6& 1.5 &    85&  1.5 &  this work \\
      19 & HAT-P-34 b &    24.3& 1.2 &     0& 14 &  this work \\
      20 & HD 17156 b &    4.1& 0.3 &     10 &  5.1 & 11 \\
      21 & HD 80606 b &    1.7& 0.3 &     42&  8 &  12 \\
      22 & HD 149026 b &    7.7& 0.8 &    12&  7 & this work \\
      23 & HD 189733 b &   3.1& 0.1 &    -0.5&  0.4 &  13 \\
      24 & HD 209458 b &    4.4& 0.2 &    -5& 7& this work \\
      25 & KOI-13 b &         65&10 &    24/154&  4 &  14 \\
      26 & Kepler-8 b &     8.9& 1.0 &     5&  7 &  this work  \\
      27 & Kepler-17 b &     4.7& 1.0 &     \multicolumn{2}{c}{$<15$} &  15  \\
      28 & TrES-1 b &         1.1& 0.3 &    30& 21 &  16 \\
      29 & TrES-2 b &         2& 1 &    -9& 12 &  17 \\
      30 & TrES-4 b &     8.5& 1.2 &    6.3&  4.7 & 18 \\
      31 & WASP-1 b &     0.7& 1.4 &   -59& 99 & -26 \\
      32 & WASP-3 b &    14.1& 1.5 &       \multicolumn{2}{c}{$3.3^{+2.5}_{-4.4}$}   &  20 \\
      33 & WASP-4 b &     2.1& 0.4 &     \multicolumn{2}{c}{$-1^{+14}_{-12}$}  & 21,22 \\
      34 & WASP-5 b &     3.2& 0.3 &     \multicolumn{2}{c}{$12.1^{+8}_{-10}$}  &  22\\
      35 & WASP-6 b &     1.6& 0.3 &    \multicolumn{2}{c}{$-11^{+18}_{-14}$}  &  23 \\
      36 & WASP-7 b &    14& 2 &    86&  8 &  24 \\
      37 & WASP-8 b &    20& 0.6 &   \multicolumn{2}{c}{$-123^{+3.4}_{-4.4}$} & 25 \\
      38 & WASP-12 b &    \multicolumn{2}{c}{$1.6^{+0.8}_{-0.4}$} &    \multicolumn{2}{c}{$59^{+15}_{-20}$} &  this work  \\
      39 & WASP-14 b &     2.8& 0.6 &   -33.1&  7.4 &  26 \\
      40 & WASP-15 b &     4.3& 0.4 &  \multicolumn{2}{c}{$-139.6^{+4.3}_{-5.2}$}  & 22 \\
      41 & WASP-16 b &     3.2& 0.9 &   \multicolumn{2}{c}{$11^{+26}_{-19}$} &  this work  \\
      42 & WASP-17 b &     9.9& 0.5 &  \multicolumn{2}{c}{$-148.7^{+7.7}_{-6.7}$}   & 27 \\
      43 & WASP-18 b &    11.2& 0.6 &    13&  7 & this work \\
      44 & WASP-19 b &     4.4& 0.9 &    15& 11 &  this work \\
      45 & WASP-22 b &     4.4& 0.3 &    22& 16 & 28 \\
      46 & WASP-24 b &     7& 0.9 &    -4.7&  4 &  29 \\
      47 & WASP-25 b &     7& 0.9 &    14.6&  6.7 & 30 \\
      48 & WASP-26 b &     2.2& 0.7 &      \multicolumn{2}{c}{$-34^{+36}_{-26}$} &    this work \\
      49 & WASP-31 b &   6.8& 0.6 &     -6& 6 & this work  \\
      50 & WASP-33 b &    86& 1 &  -107.7&  1.6 & 31 \\
      51 & WASP-38 b &     8.6& 0.4 &   \multicolumn{2}{c}{$15^{+33}_{-43}$}  & 29 \\
      52 & XO-3 b &    17.0& 1.2 &    37.3&  3.0 & 32 \\
      53 & XO-4 b &     8.8& 0.5 &   -46.7&  8.1 &  33 \\
      \noalign{\smallskip}
      \tableline
      \noalign{\smallskip}
      \noalign{\smallskip}
    \end{tabular}
    \tablerefs{ (1) \cite{gillion2010}; (2) \cite{triaud2009}; (3)
      \cite{deleuil2008}; (4) \cite{hebrard2011}; (5) \cite{johnson2008};  (6) \cite{winn2011};
      (7) \cite{moutou2011}; (8) \cite{winn2010c}; (9) \cite{winn2010e} (10) \cite{Johnson2011};
      (11) \cite{narita2009b}; (12) \cite{hebrard2010}; (13)
      \cite{cameron2010}; (14) \cite{barnes2011}; (15)
      \cite{desert2011}; (16) \cite{narita2007}; (17) \cite{winn2008}; (18) \cite{narita2010};
      (19) \cite{albrecht2011b}; (20) \cite{tripathi2010}; (21)
      \cite{sanchis2011b}; (22) \cite{triaud2010}; (23)
      \cite{gillion2009}; (24) \cite{albrecht2012}; (25)
      \cite{queloz2010}; (26) \cite{johnson2009}; (27)
      \cite{anderson2011b}; (28) \cite{anderson2011}; (29)
      \cite{simpson2011}; (30)  \cite{brown2012}; (31)\cite{cameron2010}; (32)
      \cite{hirano2011c} (33) \cite{narita2010}
    }
  \end{center}
\end{table}

\section{Discussion}
\label{sec:discussion}

We now use the measurements of projected obliquities obtained in the
last section together with other measurements taken from the
literature to learn more about the evolution of these
systems.\footnote{We present here all measurements with the sign
  convection form \cite{ohta2005} and use the symbol $\lambda$.  Often
  researchers use the symbol $\beta$ and the sign convention from
  \cite{hosokawa1953}, where $\lambda = -\beta$.}

In \ref{sec:remarks} we first remark on some specific cases from the
literature. In Section \ref{sec:tides} we revisit the evidence for an
correlation between the degree of alignment and the effective stellar
temperature discovered by \cite{winn2010}. We then discuss further
evidence of tidal interaction, based on correlations between the
planet-to-star mass ratio and $\lambda$, and the dependence of
$\lambda$ on the orbital distance between the two bodies. In Section
\ref{sec:tides} we sort the systems according to a theoretical tidal
timescale, as a test of whether tides have been important in altering
the obliquities of these systems. We also discuss the implications for
the origin of hot Jupiters and the strengths and weaknesses of our
interpretation.

\subsection{Remarks on the previous literature}
\label{sec:remarks}

\paragraph{CoRoT--1} The RM effect was measured by \cite{pont2010}. We
do not include this measurement in our subsequent discussion because
of the low SNR of the detection and because it was only possible to
obtain a few out-of-transit observations. The authors caution that
systematic uncertainties could cause the actual errors in the
measurement of the projected obliquity ($\lambda=77\pm11^{\circ}$) to
be larger than the statistical uncertainty. We note that if the value
for $\lambda$ from this study were taken at face value, it would
constitute an exception to the pattern presented below. For this
reason it would be an important system to reobserve. It is a
challenging target because of the faint and slowly-rotating host star.

\paragraph{CoRoT--11} \cite{gandolfi2010} obtained RVs during a
planetary transit in this system. As only the first half of the
transit was observed, and only with a low SNR, they could only
conclude that the orbit was prograde. In that sense this is a similar
case to HAT-P-16 presented here.  We decided to exclude CoRoT-11 from
subsequent discussion as we did with our result on HAT-P-16.

\paragraph{CoRoT--19} This system has an F9V star and a Jupiter-mass
planet on a $3.9$-day orbit. The RM effect was measured by
\cite{guenther2012}. They found $\lambda=-52^{+27}_{-22}\,^{\circ}$, a
misalignment between stellar and orbital axes. However the RM effect
was only detected at the 2.3$\sigma$ level, and no post-egress data
were obtained. We omit this measurement in the subsequent discussion.

\paragraph{KOI--13} \cite{szabo2011} detected a slight asymmetry in
the transit light curve and attributed the asymmetry to a misalignment
of the planet's orbit relative to the stellar rotation. The host star
is a fast rotator, leading to a lower surface gravity and surface
brightness around the equator compared to the poles. By modeling this
effect, \cite{barnes2011} found $|\lambda|$ to be either
$24\pm4^{\circ}$ or $156\pm4^{\circ}$.  Either choice represents a
substantial misalignment and would lead to similar conclusions in the
subsequent discussion.  For simplicity of presentation in the plots to
follow, we arbitrarily adopt the lower value. \cite{barnes2011} also
calculated the stellar inclination along the line of sight, which we
do not use here, since this information is not available for most of
the other systems. We adopt the mass for the secondary estimated by
\cite{mislis2012}, which is also consistent with work by
\cite{shporer2011c}. Both estimates were based on the photometric
orbit. We further adopt the age estimate of ($710^{+180}_{-150}$~Myr)
by \cite{szabo2011}.

\paragraph{WASP-23} We omit the measurement by
\citep{triaud2011b}. Because of a low impact parameter, the only
conclusion that could be drawn is that orbit is prograde.

\paragraph{XO--2} We omit the measurement by \citep{narita2011}. They
found $\lambda=10\pm72^{\circ}$, a prograde orbit but with a very
large uncertainty, similar to our result for HAT-P-16.

\subsection{Relevant system properties}
\label{sec:tides}

\begin{figure*}
  \begin{center}
  \includegraphics{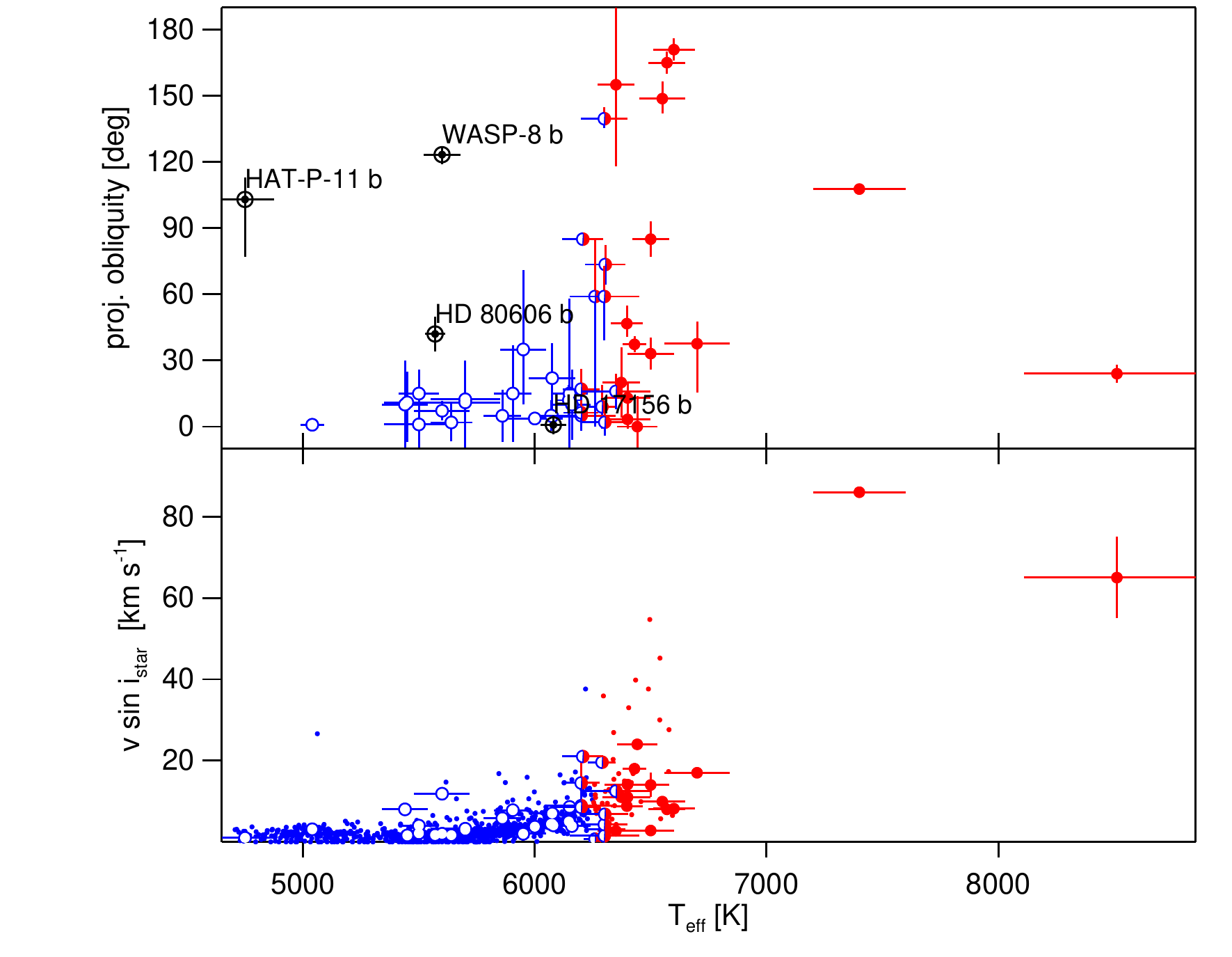}
    \caption {\label{fig:proj_obli} {\bf Projected obliquities and
        projected stellar rotation speeds as a function of the stellar
        effective temperature.}  {\it Upper panel:} Measurements of
      projected obliquities as a function of the effective temperature
      of the host star. Stars which have temperatures higher then
      $6250$~K are shown with red filled symbols. Blue open symbols
      show stars with temperatures lower then $6250$~K. Stars which
      measured effective temperature include $6250$~K in their
      1-$\sigma$ interval are shown by split symbols. Systems which
      harbor planets with mass $<0.2$~$M_{\rm Jup}$ or have an orbital
      period more then $7$~days are marked by a black filled circle
       with a ring.
      {\it Lower panel:} Projected stellar rotation speeds $v \sin
      i_{\star}$ of the stars in our sample. In addition $v \sin
      i_{\star}$ measurements of stars in the catalogue by
      \cite{valenti2005} are shown as small dots.}
  \end{center}
\end{figure*}

\paragraph{Effective temperature}

\cite{winn2010} noted that for hot Jupiters, nonzero values of
$\lambda$ tended to be associated with hot stars. For effective
temperatures $\gtrsim6250$~K the obliquities have a broad
distribution, while for lower temperatures the measurements are
consistent with low obliquities. The only exceptions were two systems
with significantly longer orbital periods than the rest.
\cite{schlaufman2010} independently found that hot stars tended to be
misaligned with the orbits of hot Jupiters by comparing the measured
value of $v\sin i_{\star}$ with the expected value for $v$, for a star
of the given mass and age. While this approach has the virtue of
requiring less intensive observations, it does rely heavily on
accurate measurement of $v\sin i_{\star}$, which is problematic for
slowly-rotating stars, and assumes that $v$ was not affected by any
tidal influence of the close-in planet.

\begin{figure*}
  \begin{center}
    \includegraphics{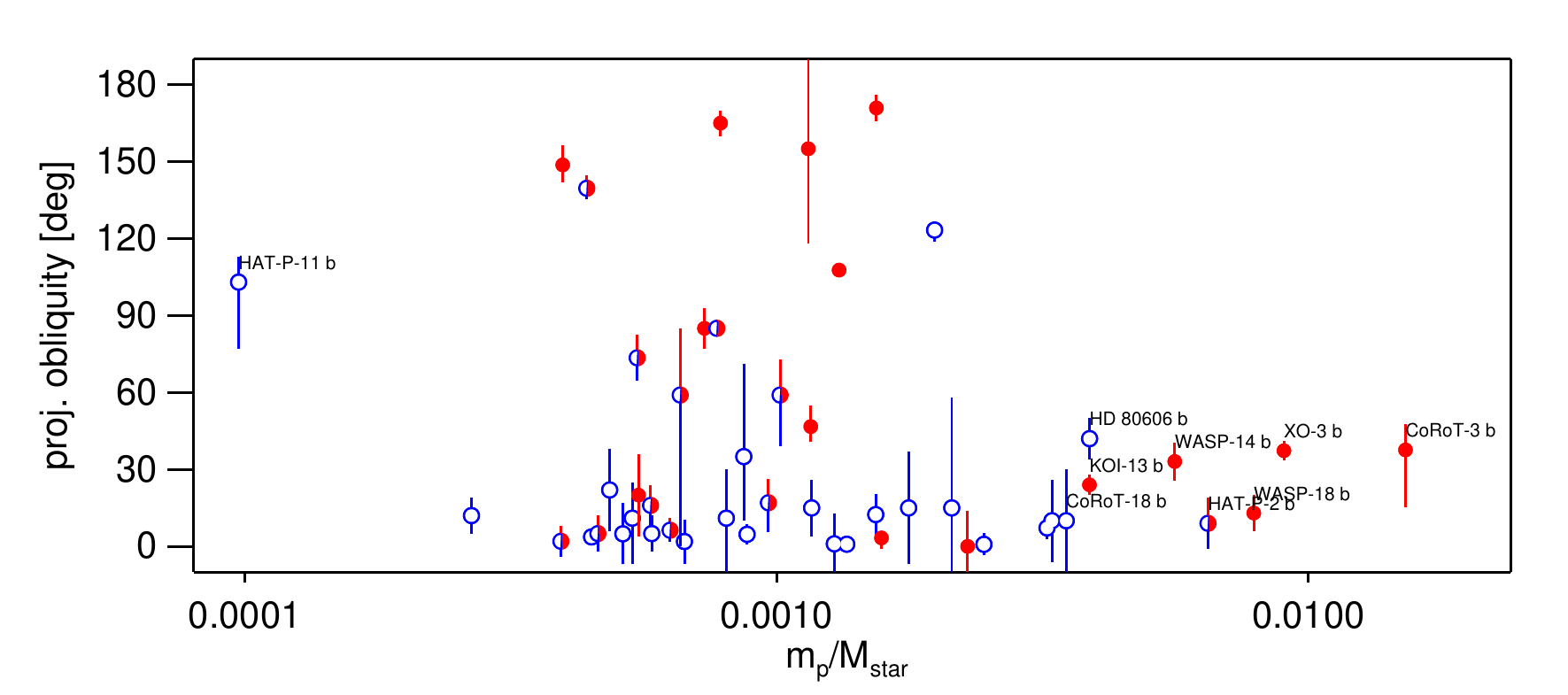}
    \caption {\label{fig:mass_ratio} {\bf Projected obliquities as
        function of the planetary to stellar mass ratio}. The same
      symbols as in Figure~\ref{fig:proj_obli} are used. With larger
      mass of the companion the degree of misalignment decreases,
      while for some of the more massive planets a small but
      significant obliquity is detected. These planets orbit stars
      with radiative envelopes, reducing the effectiveness with which
      tidal energy can be dissipated. Names of systems with particular
      small or large mass ratios are indicated.}
  \end{center}
\end{figure*}

With the new measurements presented in this paper and with
measurements by others over the last two years, the number of systems
with measured projected obliquities is now up to 53. This is more than
twice the number that was available to
\cite{winn2010}. Table~\ref{tab:proj_obli} shows the measured
projected obliquities of all systems used in this study.\footnote{See
  also {\tt exoplanets.org, exoplanet.eu,
    www.astro. keele.ac.uk/jkt/tepcat}, and {\tt
    www.aip.de/People/rheller} for listing of obliquity measurements.}
The increased number in measurements enables a stringent test of the
proposed pattern, as well as a more in depth analysis and comparison to
other system parameters.

It should be noted that in almost all cases, the measurements are of 
the projected obliquity, and not the true obliquity. For true
obliquities smaller then $90^{\circ}$ the projected obliquity is
usually {\it smaller} than the true obliquity, while for true
obliquities $>$$90^{\circ}$ the projected obliquity is usually {\it
  larger} than the true obliquity.  This factor complicates any
detailed comparison between the measurements and the theoretical
expectations. For simplicity we have chosen to work with projected
obliquities, rather than attempting any deprojection scheme \citep{fabrycky2009,morton2011}.

The upper panel of Figure~\ref{fig:proj_obli} shows the projected
obliquities plotted as a function of the effective temperature of the
host star.  Apparently, for these systems, the trend observed by
\cite{winn2010} still holds. There are three apparent exceptions to
the rule: HAT-P-11, WASP-8, and HD\,80606. These systems are special
in other ways too, by virtue of having either an unusually low planet
mass or an unusually long orbital period.  They represent three of the
four systems for which the orbital period is greater than 7~days or
the planet has a mass lower than $0.2$~$M_{\rm Jup}$. In this sense
they least resemble the typical ``hot Jupiter''. We will discuss these
important cases in the following paragraphs.

The explanation for the relationship between $T_{\rm eff}$ and
$\lambda$ could fall into one of two categories: (i) the formation and
evolution of hot Jupiters is different for hot stars than for cool
stars, which for some reason results in higher obliquities in the hot
stars.  (ii) The distribution of obliquities is originally broad for
both hot stars and cool stars, but they evolve differently depending
on a parameter closely associated with $T_{\rm eff}$. \cite{winn2010}
suggested that the second scenario is more likely and that the factor
associated with temperature is the rate of tidal dissipation due to
the tide raised by the planet.

The reason for this suspicion was that $T_{\rm eff} \approx 6250$~K is
not an arbitrary temperature, but rather represents an approximate
boundary over which the internal structure of a main-sequence star
changes substantially.  Stars hotter than this level have very thin or
absent convective envelopes, with the mass of the envelope dropping
below about $0.002~M_{\odot}$ at 6250~K \citep{pinsonneault2001}. (For
the Sun, the mass of the convective envelope is around
$0.02~M_{\odot}$).

Independently of this theoretical expectation, there is dramatic
empirical evidence for a transition in stellar properties across the
6250~K divide: hotter stars are observed to rotate more rapidly.  In
the lower panel of Figure~\ref{fig:proj_obli}, we plot the projected
rotation speeds of a sample of $\sim1000$ stars from the catalogue by
\cite{valenti2005}. The projected rotational speed $v \sin i_{\star}$
increases rapidly around 6250~K. For F0 stars, the rotation speed can
approach $200$~km\,s$^{-1}$. It is thought that stellar rotation
together with the convection in the envelope create a magnetic field
coupling to the ionized stellar wind far beyond the stellar radius and
thereby transport angular momentum away from the stellar rotation
\citep[see, e.g., ][for further discussion.]{barnes2003} Presumably
this magnetic braking is less effective for stars without convective
envelopes, leading to the observed rapid increase in stellar rotation
speeds towards earlier spectral type. Judging from
Figure~\ref{fig:proj_obli}, the transition from low obliquity to high
obliquity seems to be linked empirically to this transition from
slowly-rotating to rapidly-rotating stars.

The presence of a convective envelope is also expected to change the
rate of dissipation of the energy in tidal oscillations. Energy
contained in tidal bulges is thought to be more effectively dissipated
by turbulent eddies in convective envelopes than by any mechanism
acting in a radiative envelope. See \cite{zahn2008} for a review on
the theory of tidal interactions. \cite{torres2010} and
\cite{mazeh2008} review the evidence for tidal interactions in
double-star systems, and provide more access points to the literature.

If tidal evolution is responsible for the difference in stellar
obliquities between cool and hot stars, then there should also be a
correlation between the mass ratio of the star and planet and the
degree of alignment. The higher the mass of the planet ($m_{\rm p}$)
relative to the mass of the star ($M_{\star}$), the faster its tides
can align the stellar and orbital angular momentum vectors.
Furthermore, one would expect an inverse correlation between the
scaled semi-major axis ($R_{\star}/a$) and the obliquity. A more
distant companion will raise smaller tides. We will next investigate
whether such correlations exist. We will also investigate if the age
of the systems is an important factor in setting the degree of
alignment in these systems.

\paragraph{Mass Ratio}

In Figure~\ref{fig:mass_ratio} the measured projected obliquities are
plotted as a function of the mass ratio between planet and host
star. Higher obliquities are measured for systems in which the mass of
the planet is relatively small. This is what would be expected if
tides are responsible for the obliquity distribution. Massive planets
raise stronger tides. This trend was observed before
\cite[e.g.][]{hebrard2011b}, though at that time it was not
interpreted in terms of tidal interaction.

The interpretation is not clear from this comparison alone, though.
Note for example that the most massive planets are found around hot
stars, which should have weaker dissipation that counteracts the
effect of the more massive planet to at least some degree.  In a
subsequent section we will attempt to take both these effects (and
that of orbital distance) into account at once.

\paragraph{Distance dependence}

\begin{figure*}
  \begin{center}
    \includegraphics{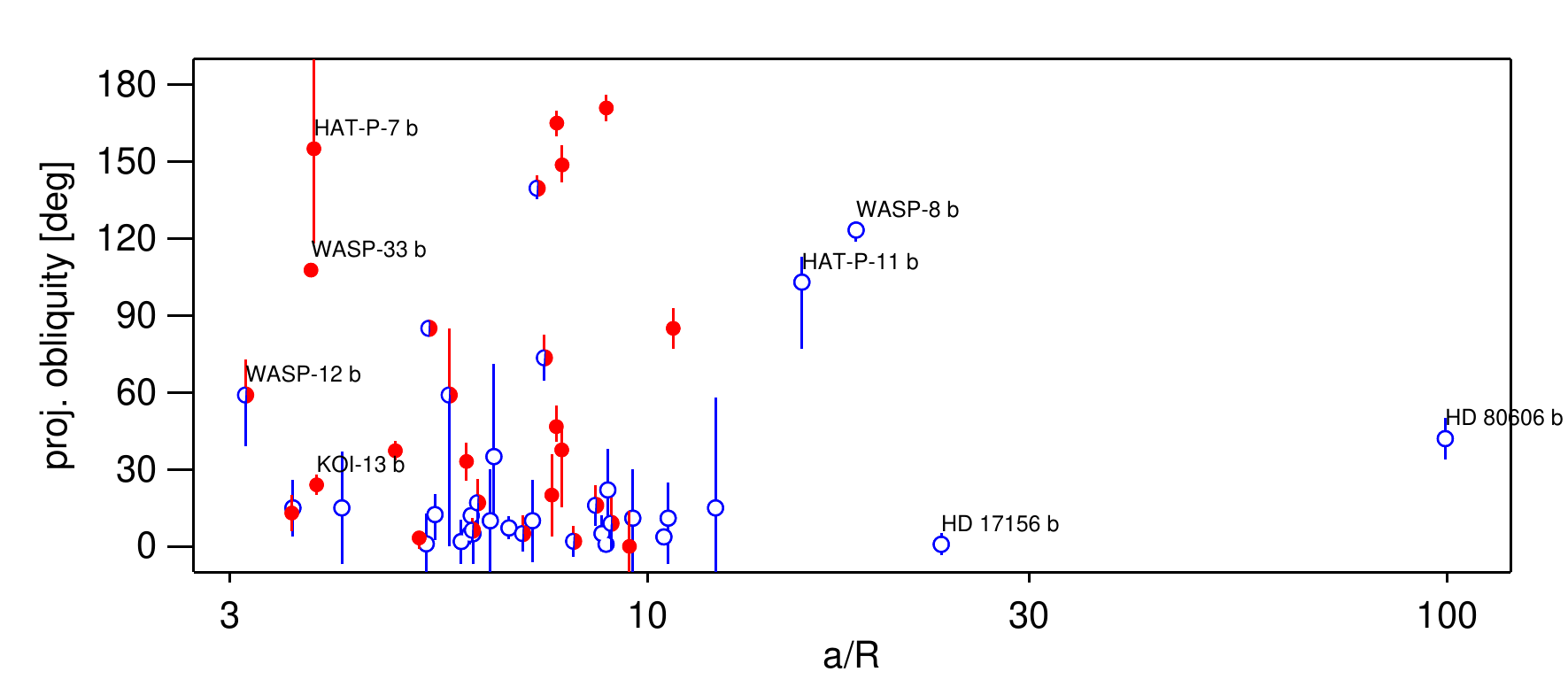}
    \caption {\label{fig:ar} {\bf Dependence of $\lambda$ on the
        scaled orbital distance.} The same as Figure~\ref{fig:mass_ratio}, but
      now the measurements of the projected obliquity are plotted as a
      function of the semi-major axis divided by the stellar
      radius. While a trend with distance can be observed for planets
      around cooler stars, such a trend seems to be absent for planets
      around hotter stars. This might be due to the small range probed
      in distance by the obliquity measurements around hot
      stars. However because tidal forces are week in these stars only
      for the innermost massive planets (e.g. WASP-12\,b and
      WASP-18\,b) such a trend would be readily observable. Names of
      systems with scaled distances greater then $15$, and for
      misaligned close in hot systems are indicated.}
  \end{center}
\end{figure*}

Does the degree of alignment depend on the scaled distance
($a/R_{\star}$)?  Figure~\ref{fig:ar} gives a mixed answer to this
question. Focusing on systems with cool stars (blue open circles)
there seems to be a trend of obliquities as a function of the scaled
distance. The data suggest good alignment for all systems with
$a/R_{\star}< 10$. Three of the four systems with greater distances
have significant misalignments.

No such dependence is observed for systems with stars which are close
to $6250$~K or hotter.  However the range of distances that is spanned
by the hot-star systems is very small, less then an order of
magnitude. In contrast, the cool-star systems probe nearly two orders
of magnitude.  Note that two of the close-in, misaligned systems are
the systems with the hottest host stars (WASP-33 and KOI-13).  This
lack of alignment finds an explanation in the tidal hypothesis:
despite the tight orbits, the tidal dissipation rate may be relatively
low due to thin or non-existent convective layer.

\paragraph{Age}
\label{sec:age}

Under the tidal hypothesis, older systems should tend to be closer to
alignment than younger systems, all else being equal.  This is because
in older systems, tides have had a longer interval over which to
act. Included in ``all else being equal'' is the underlying assumption
that the arrival time of the hot Jupiter to its close-in orbit is the
same in all systems.

\cite{triaud2011} presented empirical evidence that the degree of
misalignment depends chiefly on the age of the system.  He found that
all systems in his sample with ages greater than $2.5$~Gyr are aligned
(see his Figure~2). His sample consisted only of those stars with an
estimated mass greater than $1.2$~$M_{\odot}$, since it is harder to
determine a reliable age for lower-mass systems.

\begin{figure}
  \begin{center}
    \includegraphics[width=8.cm]{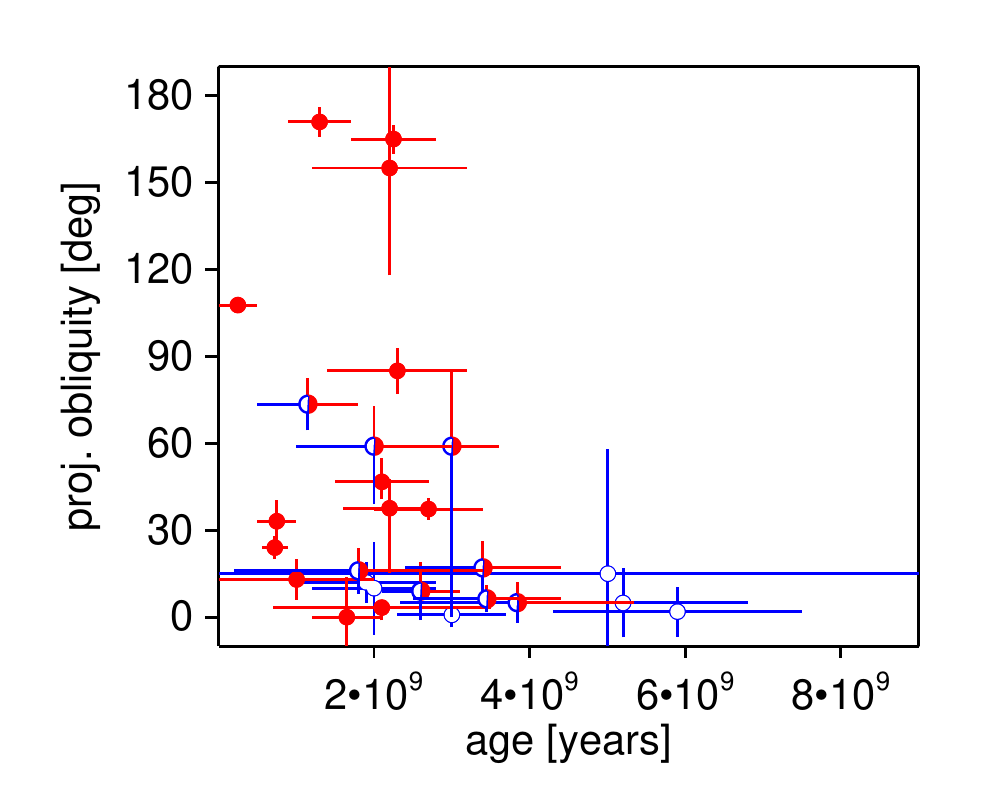}
    \caption {\label{fig:age} {\bf Projected obliquity plotted as
        function of age for stars with $M_{\star}>1.2M_{\odot}$}. This is a
      similar plot to the one presented by \cite{triaud2011}. Same
      symbols used as in Figure~\ref{fig:proj_obli}. Systems which are
      older than $\sim3$~Gyr  are cool enough to develop a convective
      envelope. This plot is therefore a relative to Figure~\ref{fig:proj_obli}.}
  \end{center}
\end{figure}

Stars with a mass of $\gsim$1.2~$M_{\odot}$ develop a significant
convective envelope during their main sequence lifetime, even if they
were too hot to have a significant convective envelope on the ``zero
age'' main sequence. In Figure~\ref{fig:age} we plot the projected
obliquities as function of stellar age for stars with $M_{\star}>1.2
M_{\odot}$. And indeed all the aligned and older systems are cool
enough to have a significant convective envelope. Figure~\ref{fig:age}
represents, therefore, a similar pattern as seen in
Figure~\ref{fig:proj_obli} with a slight shift in the variable and for
a subset of systems (only stars with $M_{\star}>1.2 M_{\odot}$). It
seems as though the development of a convective envelope with age,
rather than the age itself, might be driving the degree of alignment.

\subsection{Tidal timescale}
\label{sec:timescale}

As we have seen in the last section that the degree of alignment is
correlated with stellar temperature, the mass ratio, and possibly the
orbital distance. We now try in this section to establish a single
quantitative relationship between the degree of alignment and those
parameters.  Ideally we could calculate a theoretical alignment
timescale for each system, and compare that timescale to the estimated
age of the system.  We could then check if systems with a relatively
short timescale (fast alignment) tend to have low obliquities, and
systems with timescales comparable to the lifetime of the system (or
larger) tend to have high obliquities.

\begin{figure*}
  \begin{center}
    \includegraphics{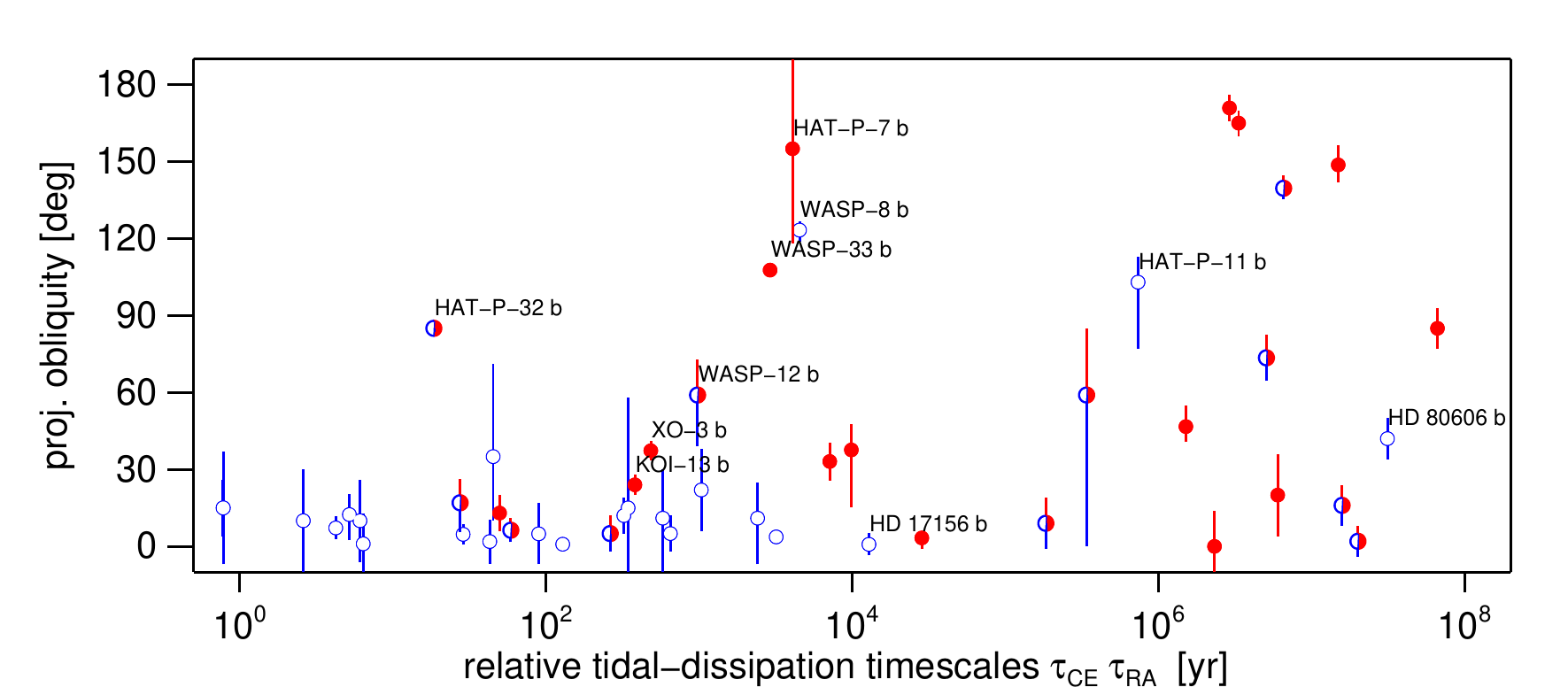}
  \caption {\label{fig:tau}  {\bf Measured projected obliquity as
      function of the alignment timescale calibrated from binary
      studies.}  The same symbols as in Figure~\ref{fig:proj_obli}
    are used. This time the projected obliquities are shown as a function
    of the characteristic timescale needed to align the stellar and orbital
    axes. We used two different equations to calculate these
    timescales. One for stars with temperatures lower than $6250$~K for
    which we assume that tidal dissipation happens due to Eddies in
    the convective envelope and one for hotter stars for which we
    assume that no convective envelope  is present and alignment is due
    to radiative damping. The coefficients for these equations have been
    calibrated with synchronization timescales in double star
    systems. Note that both timescales have been divided by $5\cdot 10^{9}$.  } 
  \end{center}
\end{figure*}

Calculating timescales needed to synchronize and align stellar
rotation is a complex task. Apart from the parameters mentioned above,
there are other parameters that would influence the time needed for
alignment. For example the total amount of angular momentum stored in
the stellar rotation, and the driving frequency of the tidal force
(i.e.\ twice the orbital frequency), are expected to be important. In
addition, the rate of dissipation is not expected to be constant over
Gyr timescales due to the contemporaneous evolution in orbital
distance and eccentricity, and due to stellar evolution.  Even worse,
the specific mechanisms for dissipating tidal energy are not
completely understood, neither for stars with radiative envelopes nor
for stars with convective envelopes. Nevertheless there are some
simple considerations we may employ to obtain approximate timescales
for alignment.

\begin{enumerate}

\item We can use the formulae provided by \cite{zahn1977} for
  synchronization. The coefficients in these formulae are difficult to
  derive from theory alone, but they can be calibrated with
  observations made in binary star systems. By observing the maximum
  orbital distance within which binary stars are observed to be
  spin-orbit synchronized, and knowing the lifetime of the stars on
  the main sequence, the relevant parameters can be estimated.  To
  apply this to our sample two different formulae are needed. One for
  cool stars which have convective envelopes (CE) and hot stars which
  have radiative envelopes (RA).  Therefore this approach has the
  virtue of empirical calibration, although the calibration is for
  star-star interactions rather than planet-star interactions, and the
  calibration is for spin synchronization rather than reorientation.
  One complication is that to apply these formulae we have to make a
  binary decision on whether a star is ``convective'' or ``radiative''
  which does not reflect the gradual thinning of the convective
  envelope with increasing stellar temperature. We choose a
  temperature of $6250$~K for this boundary.

\item Assuming that the alignment timescale due to dissipation in
  convective envelopes ($\tau_{\rm CE}$) is always shorter than the
  time needed for alignment by forces in radiative envelopes
  ($\tau_{\rm RA}$) we can try to derive a simple relationship between
  the mass contained in the convective envelope and the alignment
  timescale $\tau$. This would have the advantage that the gradual
  decrease of mass in the convective envelope can be easily
  incorporated, but we ignore here any possible additional dissipation
  mechanism in the radiative envelope which for higher temperatures
  would become important. In addition it is not obvious why
  $\tau^{-1}$ should depend linearly on the mass contained in the
  convective envelope, nor can we be completely confident in our
  estimate of the convective mass based only on the observable
  parameters of the stellar photosphere. And of course the convective
  mass is not really a constant over Gyr timescales.

\end{enumerate}

\begin{figure*}
  \begin{center}
    \includegraphics{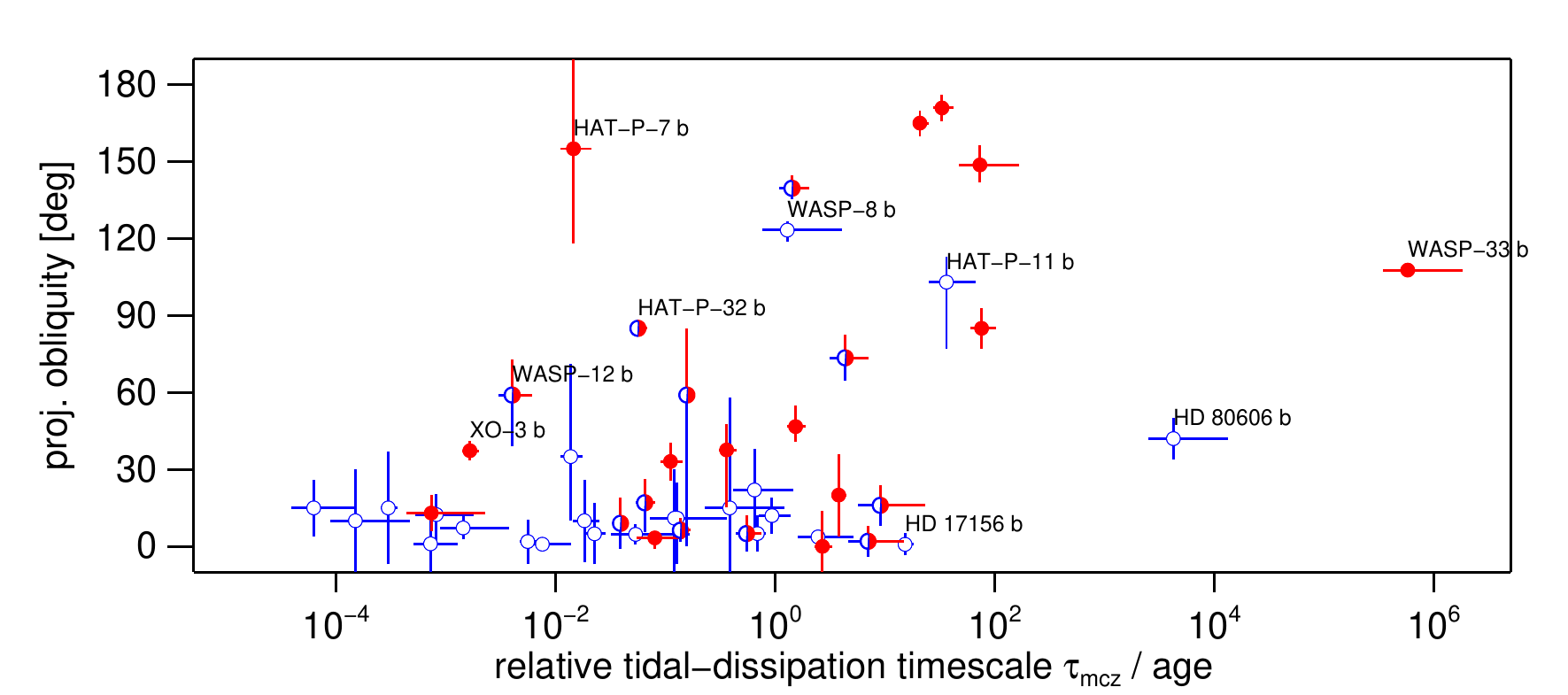}
    \caption {\label{fig:tau2} {\bf Measured projected
        obliquity as a function of the alignment timescale estimated
        from the mass of the convective envelope divided by age. }
      Similar to Figure~\ref{fig:tau}. This time however a tidal
      timescale was calculated which depends next to the mass ratio
      and the scaled distance not on a calibrated coefficient but on
      the mass in the convective envelope. Further the age estimates
      of the systems have been taken into account for this
      ranking. KOI-13 is not shown in this plot. The mass contained in
      its convective envelope is small and therefore according to
      equation~\ref{equ:tidal_time} it does practically not realign.}
  \end{center}
\end{figure*}

The simplifications made by either approach should cause us not to
expect a perfect and deterministic relationship between our
theoretical parameters and the observed obliquities [and we direct the
reader to \citet{hansen2012} for a different approach to this
comparison]. For the first approach we obtain the following
relationships between the system parameters and the convective and
radiative timescale for alignment from \cite[][and references
therein]{zahn1977},
\begin{equation}
  \label{equ:ce}
  \frac{1}{\tau_{\rm CE}} = \frac{1}{10 \cdot 10^{9} {\rm yr}} q^2 \left(\frac{a/R_\star}{40}\right)^{-6} \,{\rm   and}\, ,
\end{equation}

\begin{equation}
  \label{equ:ra}
  \frac{1}{\tau_{\rm RA}} = \frac{1}{0.25 \cdot 5 \cdot 10^{9} {\rm yr}} q^2(1+q)^{5/6}
    \left(\frac{a/R_\star}{6}\right)^{-17/2}\,\,\,.
\end{equation}
Here $q$ is the planet-to-star mass ratio. For stars with convective
envelopes, synchronization is observed for binaries out to a scaled
distance of $\sim40$ during their main-sequence life, which we set to
$10$~Gyr for all 'cool' stars.  For 'hot' stars, synchronization in
binaries is observed out to $\sim6$ times the scaled radius.  Tidal
damping is expected to be most efficient during the first quarter of
their main sequence life \citep{zahn1977} (which we set here to
$5$~Gyr). See e.g. \cite{claret1997} for a more recent comparison
between theory and observations, mainly for stars of higher mass.

In Figure~\ref{fig:tau} we show the projected obliquity versus the
characteristic timescale needed for realignment. For this we used
equation~\ref{equ:ce} for $T_{\rm eff} < 6250$~K and
equation~\ref{equ:ra} for $T_{\rm eff} \ge 6250$~K. (Both timescales
were divided by $5\cdot 10^{9}$ for normalization.) For most of the
systems for which rapid alignment is expected, low projected
obliquities are observed. For systems where tides are expected to be
too slow in aligning and synchronizing stellar rotation, a very broad
range of projected obliquities (apparently random) is observed.

We now return to the three apparent outliers from
Figure~\ref{fig:proj_obli}: HAT-P-11, WASP-8, and HD\,80606. While
these are all stars with convective envelopes, the timescales for
alignment are very long. This is because the scaled distances are
greater than $15$ (Figure~\ref{fig:ar}). In the case of HAT-P-11 there
is also the additional penalty from the relatively small planet mass
($q\approx 10^{-4}$). Thus, in this light, these three ``outliers''
are not exceptions; they have high obliquities because the tidal
timescales are very long.

There is one system which does seem to be an exception: HAT-P-32. The
rotation axis of the star nearly lays in the plane of the orbit
($\lambda=85\pm1.5^\circ$), while all other systems with a similar
tidal timescale do have projected obliquities consistent with good
alignment. Because of the obliquity near $90^{\circ}$ tides couple
only weakly to the stellar rotation \citep{lai2012}. The timescale for
alignment in this system could be longer than estimated by our simple
formula.  If this is the only reason for the high obliquity in the
HAT-P-32 system then we might expect to find more systems with short
alignment timescales and similar obliquities in Figure~\ref{fig:tau},
which do not exist in the current sample.

The measured effective temperature in HAT-P-32 is $6207\pm88$~K, and
therefore we used Equation~\ref{equ:ce}, applicable to stars with
convective envelopes. If we would have assumed an effective
temperature of $6250$~K, a value within the $1$-$\sigma$ interval of
the measurement, and used Equation~\ref{equ:ra} then we would have
obtained a $\tau_{\rm RA}$ of $1.5 \cdot 10^{5} \times 5\cdot
10^{9}$~yr instead of $\tau_{\rm CE} = 1.8\cdot 10^{1} \times 5\cdot
10^{9}$ ~yr. This illustrates the aforementioned weakness of this
first approach, that we have to make a binary decision on whether a
star is ``convective'' or ``radiative.''

Note also that KOI-13 and XO-3 are ``hot'' stars which have
significant misalignments, and yet they are found in between ``cool''
aligned systems. KOI-13 ($T_{\rm eff}=8500$~K) is hottest star in our
sample and it is questionable if we can use the same tidal timescale
for this system as for the other hot systems which are about $2000$~K
cooler (see Figure~\ref{fig:proj_obli}).

In the second approach we build upon Equation~\ref{equ:ce}. Now we do
not use any empirical calibration. For each planet-hosting star we
estimate the mass contained in the convective envelope, and assume
that the rate of energy dissipation is proportional to this convective
mass,
\begin{equation}
  \label{equ:tidal_time}
  \frac{1}{\tau} = C \cdot \frac{1}{M_{\rm cz}}  q^2 \left(\frac{R}{a}\right)^6 \,\, ,
\end{equation}
where $M_{\rm cz}$ indicates the mass contained in the outer
convective zone and $C$ is an unspecified proportionality constant
with units g~s$^{-1}$. Our estimate for $M_{\rm cz}$ is based on the
measured $T_{\rm eff}$. This ensures a gradual decrease of tidal
forces with increasing temperature. To establish the relation between
$T_{\rm eff}$ and convective mass we used the EZ-Web
tool\footnote{This tool is made available by Richard Townsend under the
  following url: {\tt http://www.astro.wisc.edu/{\tt\~{}}townsend}}
for stars with $T_{\rm eff}<7000~K$, and the data from
\cite{pinsonneault2001} for hotter stars. To create
Figure~\ref{fig:tau2} we further divided the tidal timescale
by the estimated main-sequence age, taking the uncertainty
in the age estimate into account. As the ages are not well known,
this leads to a substantial uncertainty in the positioning of a system
on the horizontal axis. On average, the hot stars are younger than the
cool stars. Therefore the main effect of the division by age is a
small shift of the hotter stars to the right side of the logarithmic
plot.

The ordering of the cool stars is not substantially changed, relative
to Figure ~\ref{fig:tau}, but there are now a few hot stars with
significant obliquities and with similar tidal timescales as some cool
aligned systems. The biggest outlier in this respect is HAT-P-7.

In summary, despite the shortcomings of our highly simplified
approaches to calculating the theoretical tidal timescale, and a few
exceptional cases, we do find support for the claim that the
obliquities in hot Jupiter systems undergo damping by tidal
dissipation.  Systems with short tidal timescales are predominantly
well-aligned, while systems with longer tidal timescales display an
apparently random obliquity distribution. The implication is that the
obliquities were once even more broadly distributed than we observe
them today.  Put differently, the ``primordial'' orbits of hot
Jupiters (the orbits that existed shortly after the planets arrived
close to the star) once had a very broad range of inclinations
relative to the stellar equatorial plane.

\subsubsection{Angular momentum problem}

As \cite{winn2010} already pointed out, there is a theoretical problem
with invoking tides in this context. The angular momentum in the
stellar rotation compared to the angular momentum in the orbit (when
the planet is close enough to significantly effect the stars rotation
via tides) is so large that to synchronize and align the star the
planet would surrender so much angular momentum that it would spiral
into the star. For nearly all systems in our sample the orbital
velocity (at periastron) is larger than stellar rotation
velocity. This causes trailing tides, and angular momentum is
transported from the orbit towards the stellar rotation, leading to
decay of the orbit \citep[e.g.][]{levrard2009}. Yet we see systems
which have aligned axes and the planets have evidently survived.

To address this problem \cite{winn2010} speculated that only the outer
layers of the star synchronize and align with the orbit. In that case
a smaller amount of the angular momentum would be transferred and the
planetary inspiral would be avoided. It seems doubtful, though, that a
separate rotation speed and rotation direction for the envelope
relative to the stellar interior could be maintained for billions of
years.

More recently \cite{lai2012} suggested that the angular momentum
problem is not as serious as it might seem.  Given the complexities of
tidal dynamics, he argued that there is no strong theoretical reason
why the timescale for realignment must equal the timescale for
synchronization, and indeed he provided a particular theoretical tidal
model in which those timescales can differ by orders of magnitude. In
his scenario the planets would first align the stellar rotation, and
only much later speed up the rotation and spiral inward.

In this respect it is interesting that the tidal timescale calibrated
via synchronization apparently sorts the systems consistently relative
to each other, but the overall timescale is too long by orders of
magnitude. As mentioned above we divided the timescales displayed in
Figure~\ref{fig:tau} by $5\cdot 10^{9}$. This could imply that
realignment happens on a shorter timescale than
synchronization. However the calibration of the synchronization
timescale was done with binary star systems having $q\approx 1$, and
the tidal mechanism itself might be different for different regimes of
its strength \citep{weinberg2012}.

One might be able to test the hypothesis of \cite{lai2012} by seeking
evidence for excess rotation in stars that are thought to have been
realigned ($\lambda \approx 0^\circ$). This could be done by measuring
the stellar rotation period or $v \sin i_{\star}$ (if one is willing
to assume $\sin i_\star$ is near unity in such systems).  If an age
estimate is also available, then one could employ the same approach as
\cite{schlaufman2010} to assess whether or not the star is rotating at
a typical rate, or if it is in the process of being spun up by the
planet. This type of analysis was pursued by \cite{pont2009}, though
not with this specific hypothesis test in mind. This analysis could be
profitably revisited now that many more systems are available for
study. There is one caveat to this approach, which is that stars
undergo very rapid spin evolution early in their lives due to
disk-locking and magnetic braking, i.e., for reasons unrelated to
planets. If hot Jupiters arrive very early in the star's history, the
realignment might happen in an epoch of rapid decrease of stellar
rotation and any memory of an increased rotation due to tides raised
by the planet might be lost.

The recent work by \cite{hansen2012} presented a calibration of the
equilibrium tide theory using the measured parameters of hot-Jupiter
systems.  While the approach taken in his work is different from ours,
he arrived at similar conclusions to those described here: tides have
shaped the obliquity distribution in these systems, and there is no
theoretical need for core-envelope decoupling.

\subsection{High obliquities: a result of dynamical interactions, or initially inclined
  disks?}
\label{sec:cause}

To interpret the finding that the host stars of hot Jupiters once had
a broad distribution of obliquities, we must answer a crucial
question. We need to know if the original obliquity is related to the
existence of the hot Jupiter, or if stars and their protoplanetary
disks are frequently misaligned for reasons unrelated to hot
Jupiters. One might expect an initially close alignment between a star
and its protoplanetary disk, as is observed in the Solar system and
has been generally assumed in the exoplanet literature.  However this
is not a foregone conclusion, and indeed several authors have recently
challenged this assumption, proposing that the Sun's low obliquity may
be an atypical case.

\cite{bate2010} proposed that a disk might become inclined with
respect to the rotation axis of the central star, as a result of the
complex accretion environment within a star cluster.  In such a dense
environment the tidal interaction with a companion star or other
nearby stars could produce chaotic perturbations in the orbits of
infalling material, with the material accreting later (destined to
become planets) having a different orientation than the material that
accreted earlier onto the star. \cite{thies2011} studied the process
of inclined infall of gas in detail and found that short period
planets on eccentric and inclined orbits can be created in this way. A
completely different mechanism for generating primordial misalignments
was proposed by \cite{lai2011b}, relying on a magnetic interaction
between a young star and the inner edge of its accretion disk.

In these scenarios, the star has a high obliquity even though the
planets may have never left the plane of the disk out of which they
have formed, and therefore the measurements of obliquities bear
information about the processes surrounding star formation rather than
planet migration. How can one distinguish between misalignment created
during the time the planet is still embedded in the disk or after the
disk dissipated?

One approach, pursued by \cite{watson2011}, is to assess the degree of
alignment between stars and their debris disks. Assuming that the
stars as well as the debris disks trace the alignment of their
predecessors one would learn about the degree of the alignment during
the final stages of planet formation. \cite{watson2011} found the
inclinations of debris disks and their stars in a sample of 9 systems
to be consistent with good alignment. They caution that in their study
only systems with $T_{\rm eff}< 6140$~K have been observed and that
misaligned system are found around hotter stars. However as we have
argued above, the found low obliquities that prevail around cool stars
may be a consequence of tidal evolution and not of the mechanism which
creates the obliquities.

Another approach is to measure the obliquities in binary star systems.
If disks would be tilted relative to stellar rotation due to close
encounters, then this could also lead to tilted rotation axes in
double star systems. There should also be a trend towards misalignment
with larger separation between the components in these
systems. Conducting such measurements and seeking evidence for such
trends is one of the goals of an ongoing observational program
entitled BANANA \citep[Binaries Are Not Always Neatly
Aligned;][]{albrecht2011}.

A more direct way to answer the question raised in the preceding
section has recently become possible, thanks to the discovery of
systems with multiple transiting planets. A number of arguments
support the idea that the orbital planes in such systems are closely
aligned; most recently \cite{fabrycky2012} used the measured transit
durations to show that the typical mutual inclinations are of order
$2^{\circ}$.  Therefore it is reasonable to assume that the multiple
planetary orbits trace the plane in which the planets formed. Any
disruptive dynamical interactions, such as those which have been
proposed to explain hot Jupiters, would likely have produced higher
mutual inclinations in the multiple-transiting systems.  Under that
assumption, RM measurements (or other measures of obliquity) in those
multiple-transiting systems would establish the angle between the
circumstellar disk and the stellar equator. If good alignment is found
to be the rule, then the high obliquities in hot Jupiter systems would
be more readily interpreted as a consequence of planet migration than
as primordial star-disk misalignments.

\section{Summary}
\label{sec:summary}

In this paper we have presented new observations of the RM effect for
14 systems harboring hot Jupiters. In addition we critically reviewed
the literature, in some cases re-analyzing data that had been obtained
previously in order to conform with our protocols. We then used these
data to show that the distribution in obliquities is consistent with
being shaped by tides raised by the hot Jupiters on the stars.  For
this we revisited the correlation between the projected obliquity and
the effective temperature discovered by \cite{winn2010}, now with a
sample of RM measurements twice as large as was then available.  We
showed that the new measurements agree with the pattern proposed by
\cite{winn2010}. With the enlarged sample we showed that obliquity in
systems with close in massive planets further depend on the mass ratio
and the distance between star and planet, in roughly the manner
expected if tides are responsible for the low obliquities.

Motivated by these results we then devised two different parameters
that represent, at least crudely, the theoretical tidal
timescales. This showed that systems which are expected to align fast
are all showing projections of the obliquities which are consistent
with good alignment. In contrast, systems for which tidal interaction
is expected to be weak, due to the stellar structure, distance, or
mass ratio, show a nearly random distribution of projected
obliquities. Our interpretation is that stars with hot Jupiters once
had a very broad range of obliquities. It is tempting to argue further
that the large obliquities originate from the same process that
produces hot Jupiters, thereby favoring explanations involving
dynamical scattering or the Kozai effect, and disfavoring the gradual
inspiral due to torques in a protoplanetary disk.  However, more
observations are needed to check on the possibility that stars and
their disks are frequently misaligned for reasons unrelated to hot
Jupiters. Among these observations are the extension of RM studies to
planets other than hot Jupiters, and measurements of obliquities in
binary star systems and in systems with multiple transiting planets.

\acknowledgments 

The authors are grateful to Nevin Weinberg, Dan Fabrycky, Smadar Naoz,
Amaury Triaud, and Ren\'{e} Heller for comments on the manuscript. Work by S.A.\ and
J.N.W.\ was supported by NASA Origins award NNX09AB33G and NSF grant
no.\ 1108595. G.B.\ and J.H.\ acknowledge the support from grants NSF
AST-1108686 and NASA NNX09AB29G. T.H.\ is supported by Japan Society
for Promotion of Science (JSPS) Fellowship for Research (DC1:
22-5935). This research has made use of the following web resources:
{\tt simbad.u-strasbg.fr, adswww.harvard.edu,arxiv.org} The W. M. Keck
Observatory is operated as a scientific partnership among the
California Institute of Technology, the University of California, and
the National Aeronautics and Space Administration, and was made
possible by the generous financial support of the W. M. Keck
Foundation. We extend special thanks to those of Hawaiian ancestry on
whose sacred mountain of Mauna Kea we are privileged to be guests.

\begin{table}
  \caption{Relative Radial Velocity measurements}
  \label{tab:rvs}
  \begin{center}
    \smallskip
    \begin{tabular}{l c c c c}
      \hline
      \hline
      \noalign{\smallskip}
      System & Time [BJD$_{\rm TDB}$] & RV~[m~s$^{-1}$] & Unc.~[m~s$^{-1}$] & Spectrograph\\
      \noalign{\smallskip}
      \hline
      \noalign{\smallskip}
HAT-P-6 &  $2455526.70653$  &  $     16.00$&$  4.97$ & HIRES \\
HAT-P-6 &  $2455526.71094$  &  $     35.72$&$  5.25$ & HIRES \\
HAT-P-6 &  $2455526.71507$  &  $     22.91$&$  5.13$ & HIRES \\
HAT-P-6 &  $2455526.71907$  &  $     -1.16$&$  5.01$ & HIRES \\
HAT-P-6 &  $2455526.72347$  &  $     -2.27$&$  4.68$ & HIRES \\
HAT-P-6 &  $2455526.72775$  &  $    -14.90$&$  4.63$ & HIRES \\
HAT-P-6 &  $2455526.73174$  &  $    -14.49$&$  4.50$ & HIRES \\
 \multicolumn{5}{c}{\vdots}\\
      \noalign{\smallskip}
      \hline
    \end{tabular}  
  \end{center}
\end{table}

\end{document}